\pdfoutput=1
\documentclass[11pt,a4paper]{article}
\usepackage[utf8]{inputenc}
\usepackage[T1]{fontenc}
\usepackage{lmodern}
\usepackage[margin=1in]{geometry}
\usepackage{amsmath,amssymb}
\usepackage{booktabs,array,graphicx,xcolor,microtype}
\usepackage{caption}
\usepackage{tikz}
\usetikzlibrary{arrows.meta,positioning}
\usepackage{longtable,pdflscape,float,etoolbox}
\AtBeginEnvironment{table}{\small}
\PassOptionsToPackage{hyphens}{url}
\usepackage[colorlinks=true,linkcolor=blue,citecolor=blue,urlcolor=blue]{hyperref}
\graphicspath{{./}}
\definecolor{resultcol}{HTML}{1F4E79}
\definecolor{passcol}{HTML}{1E7D32}
\definecolor{failcol}{HTML}{B00020}
\newcommand{\passresult}[1]{\textbf{\textcolor{passcol}{#1}}}
\newcommand{\failresult}[1]{\textbf{\textcolor{failcol}{#1}}}
\newcommand{\pp}{pp}

\newcommand{\keyresult}[1]{\textbf{\textcolor{resultcol}{#1}}}
\newcommand{\attack}[1]{\nolinkurl{#1}}
\newcommand{\dataseturl}{https://huggingface.co/datasets/Setloop/LLM-Attack-Suite}
\newcommand{\datasetrelease}{\dataseturl/tree/v2.0}
\newcommand{\datasetfile}{\dataseturl/blob/v2.0}
\newcommand{\PaperBPDFAuthors}{Georgios Politis, Evangelos Pappas}

\newcommand{\PaperBAvailability}{The archived evidence release is available at \url{https://huggingface.co/datasets/Setloop/LLM-Attack-Suite/tree/v2.0}. Data files require Hugging Face login and manual approval. The source repository is public.}

\hypersetup{pdftitle={LLM-Attack Suite: A Unified Red-Team Framework for Privacy Evaluation of Split and Distributed LLM Systems},pdfauthor={\PaperBPDFAuthors}}
\title{LLM-Attack Suite: A Unified Red-Team Framework for Privacy\\
Evaluation of Split and Distributed LLM Systems}
\author{Georgios Politis\\ {\small\texttt{geo.politis@gmail.com}} \and Setloop.io \and
Evangelos Pappas\\ {\small\texttt{evanz.pappas@gmail.com}}}
\date{}
\begin{document}

\maketitle

\begin{abstract}
Split language models keep input text on a client but send intermediate
representations to a server. Attacks can reconstruct text from these messages.
Comparing privacy evaluations is difficult when attackers see different
messages, know different model weights or use different datasets and recovery
measures. We present LLM-Attack Suite, a framework that records these choices and
lets researchers check, compare and extend an evaluation. Compatible attacks
reuse recorded messages, and their predictions are saved for independent
scoring. The scoring rule requires a linked undefended reference before a
defence can pass. Reports also identify attacks that remain untested. In a
controlled GPT-2 experiment, adding gradients returned during training increased token
recovery from 94.20\% to 97.38\% and exact-document recovery from 13.71\% to
37.77\%. The 95\% interval for the token gain lay above the
two-percentage-point threshold fixed before testing. This experiment is shared with a companion study; here it tests
how the evaluation responds to a change in attacker access. Further studies
measure how corpus choice and compression defences affect recovery, compare
an adapted published attack with its reference implementation, and test the
measurement code. The framework reports implementation checks separately
from measured attack effectiveness. Each privacy claim should state the
attacker, dataset and scoring rule, report its controls and identify untested
attacks.
\end{abstract}

\section{Introduction}
\label{sec:intro}
Split learning divides a language model between a client and a server
\cite{vepakomma2018split}. The client keeps the input text and runs the first
layers. It sends the server an \emph{activation}, the intermediate
representation computed at each token position. During training, the server
returns a \emph{gradient}, which describes how changes to these
representations would affect the training loss. Both messages contain
information about the private text.

Prior attacks reconstruct text from gradients \cite{dlg,tag,lamp},
activations \cite{vec2text,sipit} or both \cite{bisr2024}. Their results depend
on which messages the attacker sees, which model weights it knows and how
much public text and computation it uses. The dataset and recovery measure
also matter. A defence can appear effective when the attack already fails
without it. An average token score can conceal exact recovery of complete
documents or sensitive spans. Researchers need these details to interpret a
privacy result and compare it with another study.

What must an evaluation record and test so that another researcher can
check its conclusion, compare it fairly and extend it? LLM-Attack Suite\footnote{The suite was previously called LLM-Attacker. Existing software identifiers and frozen evidence may retain the earlier name for reproducibility. The separate autonomous-driving framework of Mei et al.\ \cite{mei2025llmattacker} is discussed in Section~\ref{sec:related}.} records
what the attacker can observe, what it knows and the resources it uses.
Compatible attacks run on the same recorded messages. The framework requires
a linked undefended reference before a candidate defence can pass its
scoring rule, and reports missing tests explicitly. Saved predictions let
other researchers check the scores without rerunning the target model.
The suite also includes attacks on serving, applications and agents; this
paper evaluates privacy in split and distributed systems.

The experiments test why these choices matter. Matched reconstruction
comparisons measure the effect of adding gradients to the attacker's
observations. Separate studies vary the dataset and defence. Comparisons
with a published attack implementation check agreement on specified inputs,
while tests of the measurement code check scoring, controls and handling of
the saved evidence. These checks establish different properties and are
reported separately.

The contribution has three parts:
\begin{itemize}
\item \textbf{A common evaluation procedure.} We specify how to record the
model, attacker access, data and resources, and how compatible attacks reuse
the same observations.
\item \textbf{Checkable privacy results.} We combine dataset-specific
baselines, linked undefended controls and reports of missing tests, and
retain predictions and configurations for independent scoring.
\item \textbf{Evidence for the procedure.} We assess controlled
reconstruction, agreement with a published implementation and the
measurement code. The registry records the available validation evidence for
all 64 modules, including three orchestration helpers.
\end{itemize}

This methodological contribution builds on two related studies. Politis and
Pappas \cite{politis2026privacyfailure} audit a real two-node split-training
system whose forward-message privacy check missed leakage through the
returned gradient. Their study diagnoses that system's failure and tests
observation coverage and probe calibration. A companion study by the same
authors \cite{politis2026layercount} examines how attacker observations and
model knowledge affect per-token and per-document leakage. The controlled
GPT-2 experiment reported here is shared with that study and provides one
evidence base. Here we use it to assess the evaluation procedure alongside
additional corpus, training-duration, defence and implementation studies.
The broader Qwen and Mistral experiments provide further system and
implementation checks under their own stated conditions.

\section{An Explicit Evaluation Protocol}
\label{sec:framework}
\subsection{Threat model}
\label{sec:threat}
The principal attacker passively observes the messages crossing one split.
It sees activations during training and inference, and returned gradients
during training. It neither modifies the traffic nor receives the private
text, next-token labels or server layers. The capture harness computes the
loss using the labels but does not expose them to the predictor. The scorer
uses the text afterwards. This observation contract differs from an active
server that controls returned tensors, and from a server that directly holds
the loss labels.

By default, the attacker holds a \emph{public copy} of the client's layers,
the pretrained layers available for download. A stronger condition supplies
the \emph{trained copy}, including the client's fine-tuned embeddings. The
attacker also receives public calibration text. Labelled-example, key-solving,
active-server and side-channel modules declare their additional privileges
separately. Supplementary Section~\ref{supp-sec:threat} maps the observation
surfaces and their associated modules.

\subsection{A worked comparison}
\label{sec:framework-example}
Consider the experiment at a split after GPT-2's sixth layer, with the
attacker holding public client layers. We want to measure what returned
gradients add to reconstruction. The model checkpoint, held-out documents
and public calibration text used to fit the attacker stay fixed.

\begin{enumerate}
\item \textbf{Record the observations.} Run the target on the evaluation
documents and save its activations and returned gradients. Give the
activation-only attacker the activations. Give the joint attacker both
messages. Neither attacker receives the evaluation text or next-token labels.
\item \textbf{Match the reconstruction search.} Both attacks consider 32
candidate tokens per position and keep two partial reconstructions at each
search step. The joint attacker uses gradients to propose some candidates.
Shuffled-gradient and random-candidate controls test whether any gain depends
on the gradient information. Matching the search does not match total
computation; the joint attacker also ranks candidates using gradients.
\item \textbf{Score the saved predictions.} A separate scorer compares
predictions with the held-out text. Token recovery rises from 94.20\% to
97.38\%; exact-document recovery rises from 13.71\% to 37.77\%.
Exact-document recovery requires every token in a document to be correct.
Save the predictions and settings so another researcher can check both
measures.
\end{enumerate}

This comparison isolates a change in attacker observations while fixing the
stated search limits. It also shows why the recovery measure must be
reported: the gain in complete documents is much larger than the token gain
suggests. The statistical analysis and detailed attack mechanics appear in
Sections~\ref{sec:method} and~\ref{sec:results-observe}.

\subsection{Protocol, capture and replay}
\label{sec:framework-protocol}
We describe an evaluation by
\begin{equation}
\mathcal E=(T,V,O,K,B,C,M,G).
\label{eq:evaluation-protocol}
\end{equation}
Here $T$ names the target model, split, defence and checkpoint; $V$ the
attacker's position and privileges; $O$ its observations; $K$ its knowledge;
$B$ the calibration, optimisation and search budgets; $C$ the corpus,
partitions and tokenisation; $M$ the measured outcome; and $G$ the controls,
uncertainty calculation and decision rule. These components are recorded
across the pinned protocol, run configuration and result records.

This definition identifies the experimental contrast. Reading gradients
changes $O$, receiving trained layers changes $K$, and replacing Wikipedia
with code changes $C$. An attack comparison holds the remaining components
fixed and reports each attack's assigned budget. The joint-versus-activation
experiment additionally matches candidate count and beam width, so its
controls can isolate the gradient's information.

\begin{figure}[bp]
\centering
\includegraphics[width=\textwidth]{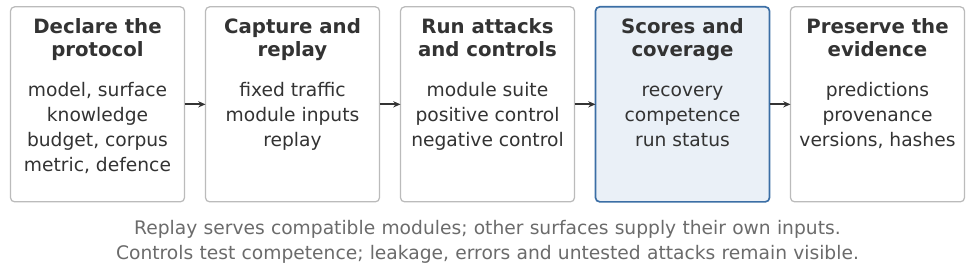}
\caption{\textbf{Researchers can check and extend an evaluation from its
recorded observations and predictions.} Compatible attacks reuse the same
messages under stated access and resource limits. Undefended references
record attack performance without a defence; other controls test the
comparison. Scores and coverage reports distinguish measured outcomes from
missing tests. Saved records link each result to its settings and source
files.}
\label{fig:framework}
\end{figure}

Capture separates target execution from attack execution
(Figure~\ref{fig:framework}). A split capture stores forward messages and,
for training, returned gradients, together with their session and ordering
metadata. Compatible attacks replay these observations against the same
checkpoint and documents. Other surfaces, such as timing or model outputs,
use their own observation providers. This arrangement supports paired
comparisons, later attack additions and independent rescoring without
retraining the victim. Regenerating predictions remains a distinct check
from rescoring saved predictions.

\subsection{Modules, coverage and evidence}
\label{sec:framework-evidence}
Each module declares its supported modes, input requirements, threat metadata
and execution function. The registry rejects duplicate names. A run writes
configuration, threat model, provenance and results to a common record,
retaining completed measurements if execution stops. A new attack can
therefore enter a frozen comparison through the same input and output
contract, with its own controls and budget.

Three outputs answer different scientific questions. The scorecard evaluates
the measured outcomes under the declared rule. The coverage report marks
applicable threats as covered, partial, unmeasured or out of scope. The
module-validation matrix records each implementation's qualification status as
reference conformance, adaptation, real-model qualification, synthetic
qualification or unmeasured capability. An execution failure is not a
negative attack result, and registration alone is not real-model validation.
The complete matrix is Supplementary Table~\ref{supp-tab:module-validation}.

The evidence procedure checks files against retained hashes, recomputes
metrics from predictions and regenerates predictions from pinned inputs.
Optional hash-chained journals reveal edits, deletion or reordering relative
to retained hashes. An unsigned chain can be recomputed, so it does not
authenticate an untrusted producer. Scientific validity instead rests on
the protocol, controls and reproducible analysis. Changes to the attack set
or scoring policy receive new versions, preserving earlier results for
comparison.

These operations serve three research tasks. Attack developers compare a new
module on existing captures. Defence researchers vary the target while
holding the evaluation plan fixed and measure both leakage and utility.
Reviewers verify and re-score the retained evidence before attempting full
regeneration.

\section{Scoring and Decision Rules}
\label{sec:gate}
The scoring rule asks how much text the strongest tested attack recovers
above a simple guess. For token reconstruction, that guess is the most
common token in the attacker's calibration text. The rule compares each
attack's upper recovery bound with this baseline, then takes the largest
difference among attacks with valid measurements. A defence can pass only
with the required repeated runs and a linked undefended reference that meets
the control rule below. A pass applies to the tested settings and attacks;
the coverage report identifies what remains untested.

Let $\hat p_a$ be attack $a$'s recovery
over $n_a$ scored tokens, and let $p_{0,a}(C)$ be the baseline's recovery on
those tokens. For applicable attacks with valid measurements, the tested
attack envelope is
\begin{equation}
L_{\mathrm{run}}(\mathcal E)=
\max_{a\in\mathcal A_{\mathrm{run}}(\mathcal E)}
100\left[U_{95}(\hat p_a,n_a)-p_{0,a}(C)\right].
\label{eq:attack-envelope}
\end{equation}
The bracketed quantity is the attack's \emph{excess}, expressed in percentage
points after multiplying by 100. $U_{95}$ is the versioned one-sided Wilson
upper bound, using $z=2.5392$. This multiplier originated in a five-percent
error budget across nine known-plaintext probe attempts and is a fixed
convention for other attacks. It is not a simultaneous confidence guarantee
over all 64 modules. Correlation among tokens also limits a literal
binomial interpretation; the gate is an operational screening convention,
not a distribution-free privacy guarantee.

A candidate defence passes when its largest excess is at most one percentage
point, its evidence includes at least three random-seed runs and a linked
undefended reference qualifies under \texttt{scorecard-policy.v3}. The
reference must cover the same attacker family and contain a result with
excess above one percentage point. The software matches report-level
references by family; explicit links are declared by the authors and require
their check. A missing or
non-qualifying control makes a candidate pass
\texttt{INSUFFICIENT\_EVIDENCE}. High measured recovery remains visible
regardless of control availability. If no valid attack ran, the envelope is
unmeasured. These rules were fixed before defence testing and retain their
named versions in the result records.

Even an attack exactly at the baseline has a positive excess at finite
sample size because its upper bound exceeds its point estimate. The record
reports this \emph{floor}. At small sample sizes it can exceed the one-point
threshold, so a failed gate must be read alongside the estimate and interval
to distinguish measured leakage from insufficient precision. The control
criterion also uses an upper bound: a sufficiently large floor can satisfy
it without high point recovery. The control's recovery, baseline and floor
require inspection; eligibility alone does not prove competence. With a fixed
protocol and existing measurements, adding an attack can only increase the
envelope or leave it unchanged. The accompanying coverage report states
what remains untested.

This token gate is distinct from the inferential tests used to compare
attacks below. Whole-document and span recovery retain their own measures.
Membership, timing and agentic outcomes do not pass through the token bound.
A verdict names its corpus, attacker set and settings, rather than extending
automatically to another protocol.

\section{Experimental Design}
\label{sec:design}
\subsection{Controlled reconstruction}
\label{sec:method}
The principal experiment uses GPT-2 124M \cite{radford2019gpt2}, split after
layer three or six, and pinned 32-token WikiText-2 passages
\cite{merity2016wikitext}. Each run uses 32 public calibration, 128 training
and 256 held-out evaluation documents. Ten independently seeded models are
fine-tuned with AdamW \cite{loshchilov2019adamw}, then frozen and attacked
with two attack seeds. The fixed snapshot supplies activations from the
client layers and gradients of the next-token loss averaged over all 256
evaluation documents. The public and trained layer conditions are evaluated
separately.

The activation-only attacker fits a ridge decoder from public activations
to token embeddings, selects 32 candidate tokens per position and uses
width-two beam search to match recomputed activations. The joint attacker
keeps 16 activation candidates and adds 16 from the negated gradient at the
previous position, filling overlaps from the activation list. The first
position retains activation candidates. Candidate count and beam width are
matched. Shuffling gradients between documents and replacing their candidates
with random tokens remove gradient information while preserving the search
structure. A backward-only heuristic provides a gradient-only comparator.
Full mechanics appear in Supplementary Section~\ref{supp-sec:method}.

The primary test was fixed in advance. At depth six with public layers, the
95\% interval for the joint-minus-activation token gain must lie above two
percentage points. The bootstrap resamples trained models and documents
independently, using 100,000 replicates and seed 93281. Forty-seven secondary
contrasts use Bonferroni-corrected bootstrap intervals. The campaign contains
3,200 training measurements and 480 explicitly reused activation-only
inference measurements, giving 3,680 records over the same held-out set.
All records and journals re-verify against their manifests.

\subsection{Additional protocols and comparators}
\label{sec:additional-design}
The corpus study applies a linear inversion decoder without beam search to
source code, templated synthetic clinical notes and WikiText-2, using public
layers at depth six. Its inference rows pool three seeds and 3,072 scored
positions per corpus. Each corpus has its own calibration-fitted baseline.

The defence-aware study optimises continuous token embeddings through the
client layers and the actual compressor. It uses the separate pilot
WikiText-2 corpus, 128 optimisation steps, three seeds and compressed widths
64, 128 and 256 across all eleven internal depths. The 198 stable-compressor
cells use public layers and training traffic. The 99 co-trained cells use
trained layers, the final compressor and held-out final-state inference
captures. These are separate observation and knowledge protocols.

We also adapt BiSR \cite{bisr2024} to the single split in our harness.
Reference conformance compares the port and archived reference in separate
processes on fresh held-out documents drawn from the WikiText-103 training
split. The subsequent paired comparison reuses the principal campaign's
targets and controls, with fresh inverters trained on public auxiliary text.
BiSR receives 20 inverter-training epochs and 32 gradient-matching plus 800
activation-matching refinement steps. Each attack retains its own fixed
budget. The comparison measures their specified configurations, rather than
an equal-compute ranking or the original two-cut BiSR protocol.

\section{Results}
\label{sec:results}
\subsection{Returned gradients increase reconstruction}
\label{sec:results-observe}
The primary test passed. At depth six with public client layers, the joint
attack recovered 97.38\% of tokens against 94.20\% for activation-only
reconstruction, a gain of 3.17 percentage points, with 95\% interval
$[2.72,3.64]$. Exact-document recovery rose from 13.71\% to 37.77\%
(Table~\ref{tab:v3primary}). All 47 secondary contrasts favoured the joint
attack after Bonferroni correction.

\begin{table}[tbp]
\centering\small
\begin{tabular}{@{}llcccc@{}}
\toprule
 & & \multicolumn{2}{c}{Token recovery (\%)} & \multicolumn{2}{c}{Exact documents (\%)}\\
\cmidrule(lr){3-4}\cmidrule(lr){5-6}
Depth & Layers held & Activation & Joint & Activation & Joint\\
\midrule
3 & Public & 97.39 & 98.29 & 43.13 & 55.51\\
3 & Trained & 97.81 & 98.72 & 51.41 & 66.13\\
6 & Public & \keyresult{94.20} & \keyresult{97.38} & \keyresult{13.71} & \keyresult{37.77}\\
6 & Trained & 94.81 & 98.27 & 20.23 & 58.55\\
\bottomrule
\end{tabular}
\caption{\textbf{A modest token gain yields substantially more complete
documents.} Each entry aggregates 5,120 document-level scores from the same 256
held-out documents across ten trained models and two attack seeds. Highlighting
identifies the prespecified primary setting.}
\label{tab:v3primary}
\end{table}

The shuffled-gradient and random-candidate controls recovered 93.91\% and
93.89\%, respectively, compared with 49.98\% for the backward-only
heuristic. The joint gain therefore depends on the gradient's information,
not simply on replacing half the candidates. Gaussian gradient noise at
0.1 times the per-document root mean square and eight-bit quantisation
left recovery at 97.31\% and 97.33\%.

\begin{figure}[tbp]
\centering
\includegraphics[width=\textwidth]{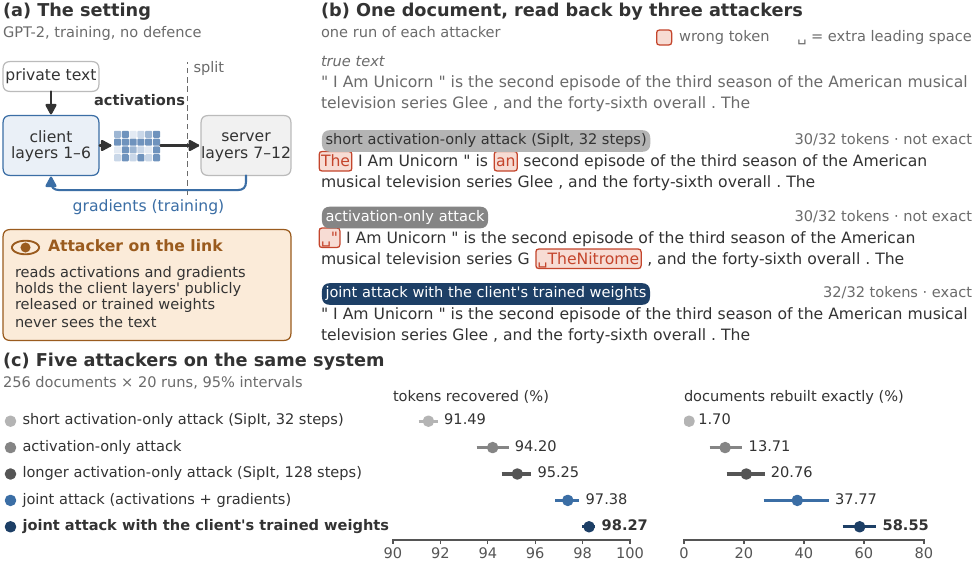}
\caption{\textbf{Attack configuration changes how much text is recovered.}
All configurations use the same depth-six GPT-2 targets and 256 held-out
WikiText-2 documents over ten trained models and two attack seeds.
(a) Observations at the split. (b) One recovered document, with errors boxed.
Among those the strongest attacker recovered exactly in that run, this one
is closest to the three attackers' average token recovery.
(c) Token and exact-document recovery with 95\% intervals from resampling
trained models and documents independently. All but the final configuration
use public layers, while the final one uses trained layers.}
\label{fig:ladder}
\end{figure}

Figure~\ref{fig:ladder} shows the wider range across attack configurations.
At the same undefended depth-six boundary, exact-document recovery varies
from 1.70\% to 58.55\%. These configurations differ in observation, knowledge
and effort. Table~\ref{tab:v3primary} supplies the matched comparison.

Defended comparisons also separate token, document and utility outcomes.
Secret mixup, which blends each activation with a private decoy from an earlier
batch, left 83.01--90.67\% token recovery while limiting exact-document
recovery to at most 1.48\%. Name-span recovery remained 47.85--66.15\%
across 716 annotated spans in 249 documents. The annotations were fixed
before the run, with AI assistance for names; these are semantic annotations,
not human gold labels. Geometric-mean perplexity rose
by 31.8\% at depth three and 26.9\% at depth six. Replica median, an
integrity defence that sends activations unchanged, kept token recovery
within two-thirds of a point of its undefended comparator. Supplementary
Section~\ref{supp-sec:results-observe} gives the defended protocols and full
span results.

\subsection{Corpus and defence change the conclusion}
\label{sec:results-text}
The same inversion attack recovered every token of the templated clinical
notes, 64.94\% of code and 63.25\% of WikiText-2
(Table~\ref{tab:domain}). Baselines also differed, from 2.64\% to 20.31\%.
All three undefended conditions failed the token gate. The clinical result
characterises regular synthetic templates. It does not measure recovery
from clinical records.

\begin{table}[tbp]
\centering\small
\begin{tabular}{@{}lrrr@{}}
\toprule
Corpus & Baseline (\%) & Recovery (\%) & Excess (pp)\\
\midrule
Source code & 20.31 & \keyresult{64.94} & 46.78\\
Synthetic clinical-style notes & 6.25 & \keyresult{100.00} & 93.75\\
WikiText-2 & 2.64 & \keyresult{63.25} & 62.79\\
\bottomrule
\end{tabular}
\caption{\textbf{Corpus choice changes both recovery and its baseline.}
Undefended GPT-2 at depth six, public layers, inference, three seeds and
3,072 pooled positions per corpus. Excess uses the Wilson upper bound,
not the recovery point estimate minus the baseline.}
\label{tab:domain}
\end{table}

Knowing the defence also matters. The validated white-box path recovered
18.85--68.59\% mean tokens across the stable-compressor depth profiles,
despite reducing each 768-dimensional activation to 64--256 numbers.
Co-trained bottlenecks showed 23.52--67.68\% recovery across their separate
final-state protocol. Each depth mean averages three widths and three seeds.
Twenty checks validated the gradient path, work accounting and capture
matching. All co-trained captures replayed exactly from their saved parts.
The full depth table and budgets are in Supplementary
Section~\ref{supp-sec:results-correct}.

Horizon and longer-context studies extend these settings without merging
their protocols. No verdict changed between 20 and 100 training steps in
the six-defence pilot battery. A 500-step extension found that a fixed
random projection suppressed the declared, projection-unaware inversion
attacker below its majority baseline, but also largely prevented useful
learning. Cross-entropy remained 7.42--7.75 against 4.39--4.44 without a
defence. On 64- and 128-token code and Wikipedia passages and 32-token
dialogue, undefended inversion recovery ranged from 76.56\% to 90.22\%.
Gaussian activation noise with $\sigma=15$ reduced it to 9.96--55.05\%.
The supplement retains the individual conditions, controls and utility
measurements rather than combining them into one attack-strength score.

\subsection{Published-attack conformance and comparison}
\label{sec:results-published}
Our BiSR port and the archived reference produced identical tokens in all
3,072 paired predictions under the single-seam contract. This establishes
implementation conformance on those inputs before comparison with other
attacks.

In the adapted experiment, refined BiSR recovered 93.21\% at depth six with
public layers, against 97.38\% for the joint attack. The paired difference
was $-4.17$ percentage points, with 95\% interval $[-4.79,-3.53]$.
With trained layers, the difference was $-0.02$ points, interval
$[-0.63,0.58]$. Across 15 prespecified secondary contrasts, ten intervals
lay below zero, one above and four crossed zero. The configurations in
Table~\ref{tab:bisr-adapted} show why the framework retains several attacks
and their budgets rather than naming a universal winner.

\begin{table}[tbp]
\centering\small
\begin{tabular}{@{}llrrrr@{}}
\toprule
Depth & Layers held & Activation & SipIt-128 & Joint & Refined BiSR\\
\midrule
3 & Public & 97.39 & 97.45 & 98.29 & 96.16\\
3 & Trained & 97.81 & 98.89 & 98.72 & 97.79\\
6 & Public & 94.20 & 95.25 & \keyresult{97.38} & \keyresult{93.21}\\
6 & Trained & 94.81 & 98.14 & 98.27 & 98.25\\
\bottomrule
\end{tabular}
\caption{\textbf{No tested configuration wins in every setting.} Token
recovery (\%) on the same 256 documents, ten trained models and two attack
seeds, with attack-specific fixed budgets. Activation, SipIt and joint
results are reused controls from the main campaign. The inverter-only stage,
conformance protocol and complete work counts appear in the supplement.}
\label{tab:bisr-adapted}
\end{table}

\subsection{Validation of the measurement procedures}
\label{sec:results-method-validation}
We selected 25 existing development-test functions before rerunning them
from the reviewed framework revision. Parameterisation yielded 37 cases
covering recovery metrics, score calculations, control eligibility,
observation boundaries, solver fidelity, resampling and evidence validation.
The design combines known-answer fixtures, independent calculations, fault
injection and metamorphic tests of prescribed relationships between outputs
under controlled changes to inputs \cite{xie2011metamorphic}.
Inputs comprised constructed sequences and arrays, small randomly initialised
models and synthetic evidence records.

All 37 selected cases matched their expected outcomes, with no failures or
skips. Changing text available only to the BiSR scorer left predictions
byte-identical. Supplying that text to the predictor was rejected. Removing
or weakening a required positive control prevented a candidate pass. In a
dependence stress case, increasing perfectly correlated document columns
from two to 500 preserved the interval $[0,1]$ instead of treating copies
as independent evidence. Intact evidence was accepted and the specified
corruptions rejected. These fixed-case results establish conformance of the
selected measurement procedures, complementing reference conformance and
real-model experiments. Supplementary Table~\ref{supp-tab:measurement-validation}
reports all seven property groups and their case counts.

\section{Cumulative Evidence and Scope}
\label{sec:limitations}
Taken together, the programme establishes consequential leakage under
several independent protocols. The findings include exact structural leakage in a deployed Qwen
split-training system, high text reconstruction in the controlled GPT-2
campaign, and further corpus-, placement-, depth- and defence-dependent
failures. The Qwen audit recovered the real/decoy partition on all 4,096
frames in its main run even though the forward-only privacy gate passed
\cite{politis2026privacyfailure}. Structural recovery and text reconstruction
remain distinct outcomes.

The broader qualification programme covers Qwen3-0.6B, Qwen3-8B,
Qwen3.5-27B, Qwen3.5-35B-A3B, Mistral-7B-Instruct and
Mistral-NeMo-12B-Instruct in diagnostic, ablation, robustness, obfuscation
and agentic roles. These campaigns record expected, negative and
privacy-breaking outcomes across multiple models and defences. They establish
that modules have been exercised under their declared contracts, alongside
the controlled GPT-2 recovery results. Supplementary
Table~\ref{supp-tab:strength-evidence} records each protocol's threat model,
module, control, model, corpus, budget, outcome, comparator and evidence tier.

The live agentic tier illustrates the importance of retaining negative
outcomes. A Qwen3-0.6B planner with reasoning disabled produced measurable
actions in all ten fixed settings and no policy violation in twenty trials,
but blocked all six sanctioned peer-request trials. This qualifies
planner/tool integration and identifies over-blocking. It does not establish
general agentic security. Synthetic qualification, reference conformance
and target-specific security measurements occupy separate tiers throughout
the module matrix.

The strongest quantitative reconstruction comparisons in this paper concern
GPT-2 124M under the stated corpora, lengths and budgets. Their main value
for the framework is that changing a protocol component changes the
conclusion, while retained captures and predictions make that change
inspectable. Corpus-specific baselines, utility measurements and matched
controls supply the interpretation. Aggregating different protocols into a
single success rate would discard it. Longer contexts, new targets and
additional attack budgets can extend the evidence through the same procedure.

All experiments used author-controlled systems. Live endpoint modes require
an explicitly supplied host, and the agentic planner accepts local model
targets only.

\section{Related Work}
\label{sec:related}
VFLAIR-LLM \cite{vflair} combines configurable split inference and fine-tuning
with attacks and defences. garak \cite{garak2024} organises security probing
through generators, probes, detectors and reports, while ML Privacy Meter
\cite{murakonda2020privacymeter} measures training-data privacy risk through
membership inference. LLM-Attack Suite concentrates on the experimental contract
for split and distributed systems, covering explicit observations and knowledge,
shared captures, linked controls, execution coverage and versioned evidence.
Our comparisons test the consequences of those choices.

Mei et al.\ \cite{mei2025llmattacker} introduced LLM-Attacker for
LLM-guided closed-loop adversarial scenario generation in autonomous
driving. Our suite was previously released under the same name.
LLM-Attack Suite provides modular attacks against LLM systems, with this
paper focusing on split and distributed privacy evaluation.

Gradient reconstruction includes DLG \cite{dlg}, TAG \cite{tag} and LAMP
\cite{lamp}, while activation inversion includes Vec2Text \cite{vec2text} and
SipIt \cite{sipit}. BiSR \cite{bisr2024} combines a learned inverter with
gradient and activation refinement for split fine-tuning. We assess its
reference conformance under the local single-seam contract and compare the
adaptation with fixed-budget alternatives. The remaining modules are local
implementations or adaptations with their evidence tiers stated separately,
not claims of reproducing every cited paper's complete protocol.

Earlier split-learning attacks demonstrated inversion, model stealing and
label inference \cite{pasquini2021unleashing,erdogan2022unsplit}. Subsequent
work examines why depth alone does not guarantee privacy
\cite{dong2025depth} and recovery during split inference \cite{gu2026pidi}.
These motivate evaluation across split depths and observation surfaces.
The supplementary catalogue records the contracts and qualification evidence
for membership, side-channel, application and agentic modules. Their endpoints
remain separate from the text-reconstruction gate.

\section{Conclusion}
\label{sec:conclusion}
LLM-Attack Suite makes privacy evaluation an explicit, reusable experiment.
Its modules share observation contracts and evidence records, its scoring
requires baselines and positive controls, and its coverage report keeps
untested threats visible. The experiments establish that attacker knowledge,
returned gradients, corpus, defence and attack configuration materially
change recovery. Reference conformance and selected measurement-validation
tests support the procedures used to measure those effects.

Researchers can use the same framework to add attacks to retained captures,
compare defence revisions and give reviewers predictions they can re-score.
Every privacy claim for a split model should therefore name its attacker,
corpus and scoring rule, report the supporting controls and leave untested
attacks labelled as untested.

\section*{Evidence and Supplementary Material}
The supplementary material appended to this article retains the full
protocols, per-condition results, 64-module validation matrix, source
identifiers and reproduction guide. Its sections, figures, tables and
equations use an S prefix; all cross-references resolve within this document.
The canonical code release is \url{https://github.com/setloop-io/LLM-Attack-Suite/tree/v2.0}.
The retained campaign evidence, measurement-validation records and executable
source snapshots are collected in \href{\datasetrelease}{Hugging Face release v2.0}.
This release reuses the unchanged scientific evidence from v1.0 and updates
branding and paper packaging only; it adds no experiments or runtime changes.
The additional 37-case validation package is at
\href{\datasetrelease/validation/measurement-20261001}{\nolinkurl{validation/measurement-20261001/}}.
The release includes a verified source archive and a Git-independent runner
so that the validation can be reproduced without the former repository history.
File verification, rescoring and prediction regeneration are documented in
Supplementary Section~\ref{supp-app:release}.
\PaperBAvailability
\label{end:main}

\section*{Acknowledgments}
AI assistance was used for restructuring, drafting, grammar checking and
editing the main text and appendices. Also final LaTeX corrections, preparation of the
publication variants, github repository housekeeping and submission documentation.
Experimental measurements and numerical results were taken from the retained study records, no new experimental results were generated
during this publication-preparation step.

\clearpage
\section*{Supplementary Material}
\setcounter{section}{0}
\setcounter{subsection}{0}
\setcounter{table}{0}
\setcounter{figure}{0}
\setcounter{equation}{0}
\renewcommand{\thesection}{S\arabic{section}}
\renewcommand{\thetable}{S\arabic{table}}
\renewcommand{\thefigure}{S\arabic{figure}}
\renewcommand{\theequation}{S\arabic{equation}}
\renewcommand{\theHsection}{supp.\arabic{section}}
\renewcommand{\theHsubsection}{supp.\arabic{section}.\arabic{subsection}}
\renewcommand{\theHsubsubsection}{supp.\arabic{section}.\arabic{subsection}.\arabic{subsubsection}}
\renewcommand{\theHtable}{supp.\arabic{table}}
\renewcommand{\theHfigure}{supp.\arabic{figure}}
\renewcommand{\theHequation}{supp.\arabic{equation}}
The following sections retain the detailed threat models, attack procedures,
experimental results, release information and complete 64-module validation
matrix behind the main text. They describe the same experiments and
artifacts, not an additional independent evidence base. Unprefixed references
identify the main text. The retained campaign and measurement-validation
artifacts are collected in tag \texttt{v2.0}, as specified in
Section~\ref{supp-app:release}.

\section{Threat Model and Setup}
\label{supp-sec:threat}

\begin{figure}[H]
\centering
\includegraphics[width=\textwidth]{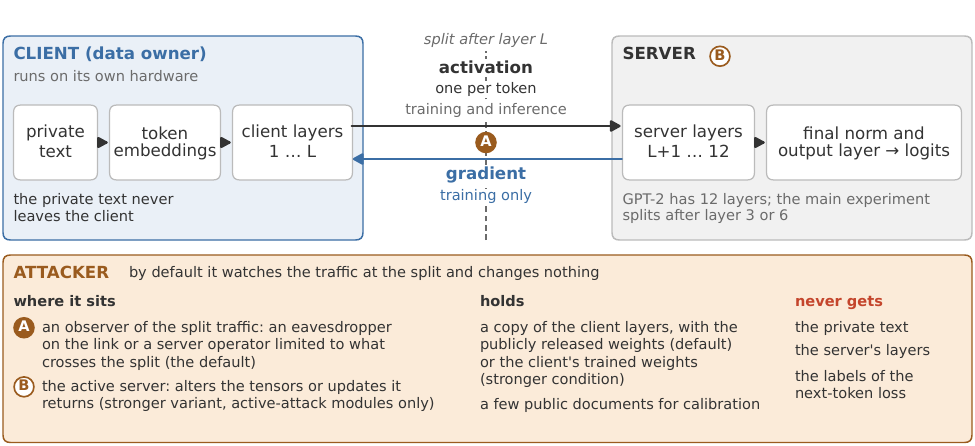}
\caption{\textbf{The client keeps the text and its first layers; the attacker reads what crosses the split.} Marker A is our attacker, a passive observer of the split traffic. Marker B is the active server, a stronger attacker used only by the active-attack modules.}
\label{supp-fig:system}
\end{figure}

Figure~\ref{supp-fig:system} shows the two parties, the traffic between them and where our attacker sits. The \emph{client} owns the text and runs the first layers, from the token and position embeddings through the block at the split depth. The \emph{server} runs the rest: the remaining blocks, the final layer norm and the output projection to vocabulary logits. In training and in inference the client sends the server an activation for each token; during training it also receives back a gradient for each activation, which is the next-token loss differentiated with respect to that activation.

Our \emph{attacker} watches this traffic at the split and alters none of it. It can be an eavesdropper on the link or an operator of the server who sees only the tensors that cross the split. It records every activation and, in training, every gradient, and it keeps its own copy of the client's layers along with a handful of public documents for calibration. The private text and the server's layers stay out of its reach, and so do the next-token labels, which are the text shifted by one position. For this reason we do not treat the whole server as our attacker, since a server that computed the loss itself would hold those labels. The code that captures the traffic in our experiments runs the server's layers and computes the loss from the labels, yet it passes neither to the attack; we use the text only to score the attacks.

A server that alters the tensors or updates it returns, to make the client leak more, is a separate and stronger attacker, the \emph{active server}; only the active-attack modules take that role.

The attacker's copy of the client's layers is by default the \emph{public copy}, the pretrained model's first layers as publicly released, which anyone can download. As a stronger condition we give it instead the \emph{trained copy}, the client's first layers after training together with the client's fine-tuned token embeddings. The client does not publish these weights; we give them to the attacker to measure what a leak of them would add, and we count the time to export them in the attack's cost.

Some modules assume a stronger attacker still, and each module states which one it assumes. A \emph{labelled} attacker also holds activations whose tokens it knows, either because it chose that text itself or because some source reveals the true token behind an observed activation; we call such a source an \emph{oracle}. A \emph{label-free} attacker has only public text and public weights. Some defences also multiply every activation by a secret key and replace the key at each \emph{epoch}, one full pass over the training data; we call this \emph{secret rotation}. A \emph{per-epoch} attacker must start again after each new key, while a \emph{cross-epoch} attacker carries information from one key to the next, for example a structure that all keys share or key material leaked in earlier epochs. The key-solving modules model these attackers; the one experiment in this paper that runs them uses no rotation (Section~\ref{supp-sec:results-text}).

The same modules run from three positions without code changes: installed next to the defence, probing a fleet of machines over a network connection, or as a relay on the client-to-server link. The relay forwards both directions unchanged and copies them into a log; once the log reaches its size limit, it stops copying and keeps forwarding.

\begin{figure}[H]
\centering
\includegraphics[width=\textwidth]{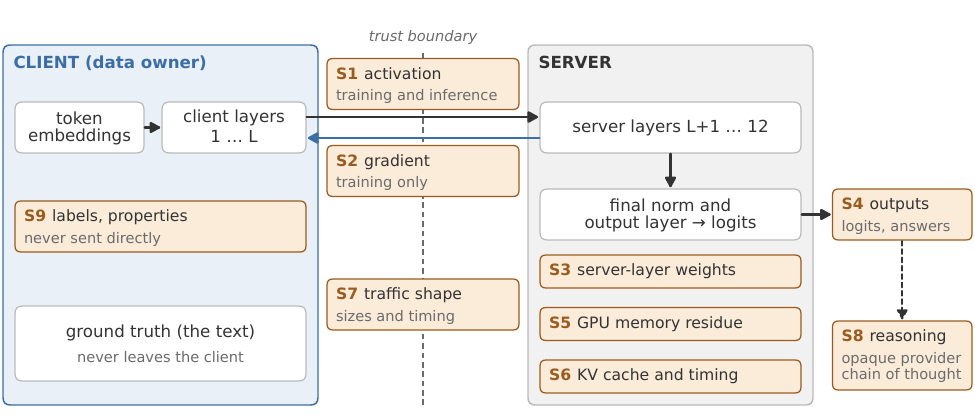}
\caption{\textbf{An attacker can observe three surfaces on the link, one on the client and five on the server side.}
The client holds the private text, the token embeddings and the client layers; the server holds the remaining layers, the final norm and the output layer. Orange boxes are the observation surfaces S1--S9 of the surface model; control-plane metadata, persistent stores and router state extend it. Table~\ref{supp-tab:surfaces} lists the attacks on each surface. Cutting the model twice, into three parts, adds a second place where activations cross; mixture-of-experts server layers leave the activation unchanged but add routing decisions as a possible leak.}
\label{supp-fig:anatomy}
\end{figure}

The activation and the gradient are the first two of the nine observation points in Figure~\ref{supp-fig:anatomy}. The other seven are the server's weights, its outputs, leftover GPU memory, caches and timing, the shape of the traffic, reasoning records returned by an API, and the labels and properties of the data. Apart from the shape of the traffic, these points sit on the server or the client, beyond what an observer at the split can see, so the modules that read them assume a stronger attacker. Table~\ref{supp-tab:surfaces} lists what crosses each point and the modules that attack it, including five points we added after reviewing the threat model.

\begin{table}[H]
\centering\small
\footnotesize
\begin{tabular}{@{}c>{\raggedright\arraybackslash}p{2.9cm}>{\raggedright\arraybackslash}p{4.2cm}>{\raggedright\arraybackslash}p{6.1cm}@{}}
\toprule
\# & Observation point & What crosses it & Attacks on it \\
\midrule
S1 & activation (forward) & the activations of each training batch or prompt &
\attack{seam-inversion}, \attack{vec2text-inversion}, the key-solving
modules for secret rotation, \attack{latent-probe},
\attack{latent-sensitivity}, \attack{wire-eval} \\
S2 & gradient (backward) & the gradients returned for those activations &
\attack{gradient-inversion} (DLG++, our multi-restart extension of deep
leakage from gradients; iDLG; and a local variant of TAG), \attack{batch-label-recon},
\attack{label-scoring}, joint reconstruction \\
S3 & server-side weights & never (held by server) &
\attack{body-extraction} (functional distillation) \\
S4 & outputs / logits / answers & API responses &
\attack{output-inversion}, \attack{trace-inversion} \\
S5 & GPU memory residue & freed buffers, local memory &
\attack{gpu-memory} (residue, scrub-check) \\
S6 & cache / timing & key-value and prefix caches, contention, acceptance patterns &
\attack{kv-timing} (with a speculative-decoding acceptance mode),
\attack{gpu-memory} (unified-memory timing mode) \\
S7 & traffic shape & packet sizes/timing on the link &
\attack{wire-shape} (via \attack{mitm-proxy} capture) \\
S8 & API reasoning records & encrypted chain-of-thought records &
\attack{reasoning-replay} \\
S9 & labels / properties & never directly &
\attack{lira-membership}, \attack{membership}, \attack{label-scoring} \\
\midrule
\multicolumn{4}{@{}l}{\emph{Extended observation points, added after our
review of the threat model:}}\\
S10 & protocol / control plane & negotiation, session binding, epochs, keys &
\attack{protocol-integrity}, \attack{key-session-abuse},
\attack{seed-disclosure} \\
S11 & persistent stores / telemetry & captures, logs, checkpoints, crash
dumps, backups & \attack{artifact-telemetry-leak}, \attack{remote-fetch} \\
S12 & mixture-of-experts router metadata & expert IDs, queue depth, capacity overflows &
\attack{router-side-channel} \\
S13 & AI artefacts / supply chain & models, tokenisers, adapters, datasets,
plugins & \attack{supply-chain}, \attack{poisoning-backdoor} \\
S14 & application context & system prompts, retrieved documents, tool schemas,
memory & \attack{prompt-injection}, \attack{hidden-context-exposure},
\attack{rag-vector-abuse} \\
\bottomrule
\end{tabular}
\caption{\textbf{Where each attack reads, and which modules read it.} What
crosses each observation point of Figure~\ref{supp-fig:anatomy} and the modules
that exploit it.}
\label{supp-tab:surfaces}
\end{table}

\section{How the Attacks Work}
\label{supp-sec:method}

\begin{figure}[H]
\centering
\includegraphics[width=\textwidth]{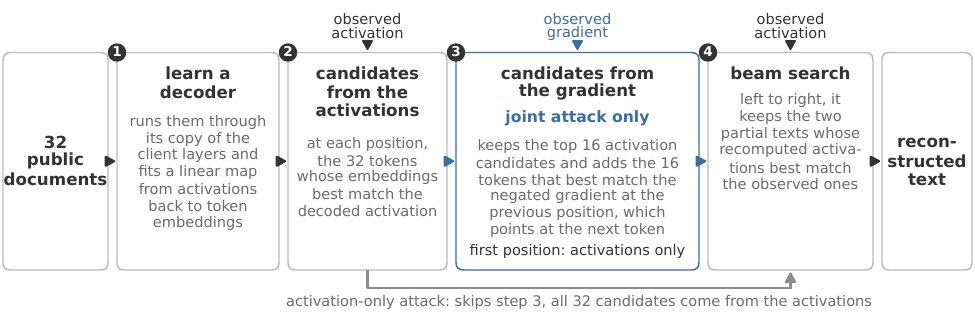}
\caption{\textbf{The joint attack takes half of its candidate tokens from the gradient and otherwise runs the same steps as the activation-only attack.}
The numbered badges correspond to the four stages described below.}
\label{supp-fig:method}
\end{figure}

The strongest reconstruction attack rebuilds a document in four steps. The
activation-only attacker uses steps one, two and four; the joint attacker
adds step three. This attack is a self-contained routine, distinct from the
registered inversion module's own joint variant, which explores candidates
differently. The four steps match the numbered badges of
Figure~\ref{supp-fig:method}.

\begin{enumerate}
\item \textbf{Learn a decoder.} The attacker sends 32 public documents
through its copy of the client's layers and records, for each token, the
activation and the token behind it. From these pairs it fits a linear map, a
ridge regression held stable by a penalty on its size, that takes an
activation back to a token embedding, the vector the model gives each word
in its vocabulary.
\item \textbf{Candidates from the activations.} At each position, the
attacker applies this map to the observed activation and lists the 32
tokens whose embeddings have the highest \emph{cosine similarity} with the
result, meaning they point most nearly in the same direction.
\item \textbf{Candidates from the gradient (joint attack only).} The
gradient comes from the next-token loss, whose target at each position is
the document's next token, so the gradient at one position carries a
signal about the token that follows it. The joint attacker keeps the top 16
candidates from the activation and adds the 16 tokens whose embeddings have
the highest cosine similarity with the negated gradient at the previous
position, filling any overlap from the activation list. No earlier
position predicts the first token, so the first position keeps its
activation candidates.
\item \textbf{Beam search.} A width-two search reads the text left to right,
extending its two leading guesses with each candidate, pushing them back
through the client's layers, and retaining the two whose recomputed
activations lie closest to what crossed the split. The two attackers try
equally many partial texts and part ways only on their candidate lists.
\end{enumerate}

In the main experiment the attacker reads one fixed snapshot of this
traffic rather than live training traffic. We fine-tune the whole model, freeze
it, split it, and pass the 256 evaluation documents through it once. The
activations come from the frozen client layers, and the gradients from the
next-token loss averaged over all 256 documents.

The \emph{shuffled-gradient} control strips the gradient of its information
by giving the joint attacker another document's gradient in place of the
right one. The \emph{random-candidate} control retains the activation half of
the list and draws the rest uniformly from the vocabulary, so no gradient
enters. The \emph{backward-only} heuristic uses the gradient alone: at each
position it picks the token with the highest cosine similarity to the
negated gradient at the previous position, with no activations and no beam
search. We also run SipIt \cite{sipit}, which rebuilds the input
one position at a time: for each position it searches for the token whose
activation, given the prefix already recovered, matches the observed one,
with a budget of steps per position. We implemented the published algorithm
ourselves; when the budget runs out it returns the best token it has tried,
so its guarantee of exact recovery no longer holds.

The experiments also use a \emph{white-box}
attack, which knows the defence and searches for continuous token embeddings that,
after passing through the client's layers and the defence, reproduce the
observed message, then reads off the nearest token at each position.
The \emph{inversion attack}, a form of vanilla model inversion, fits the
same linear map from activations to embeddings as step one, on its own
calibration text passed through the public layers, and reads off the
nearest token without a beam search. \emph{BiSR}, or bidirectional semi-white-box
reconstruction \cite{bisr2024}, works in two stages: a learned inverter
first guesses the text from the forward activations, and an optimisation
stage then refines the guess. In our port, the inverter (which the code
names SIP) is a small recurrent network trained on public text passed
through the public layers, and the refinement (which the code names bf)
first matches the observed gradient and then matches the observed
activation. Table~\ref{supp-tab:signals} lists each signal, when it crosses, how
an attack uses it and what it reveals.

\begin{table}[H]
\centering\small
\renewcommand{\arraystretch}{1.15}
\begin{tabular}{@{}>{\raggedright\arraybackslash}p{0.2\textwidth}>{\raggedright\arraybackslash}p{0.13\textwidth}>{\raggedright\arraybackslash}p{0.33\textwidth}>{\raggedright\arraybackslash}p{0.24\textwidth}@{}}
\toprule
Signal & When it crosses & How the attack uses it & What it reveals \\
\midrule
Activation, one vector per token & Training and inference & The learned
decoder turns each activation into a short list of likely tokens; a beam
search keeps the guesses whose recomputed activations match. SipIt and
the inversion attack also read it. & Token, whole-document and span
recovery \\
\addlinespace
Gradient of the next-token loss & Training only & Each gradient points at
the token that follows it, so it supplies half of the candidates at each
position. & What the gradient adds to the same scores \\
\addlinespace
Size of the returned gradient, per document & Training & The summed
gradient size scores whether a document contains a given token. &
Properties of the text, as an area under the ROC curve \\
\midrule
\emph{Controls:} another document's gradient; random tokens in place of
the gradient's suggestions & Training & The same attack, with the
gradient's information removed & Whether the gain comes from the gradient \\
\bottomrule
\end{tabular}
\caption{\textbf{What an observer at the split reads, and what each signal
gives up.} The controls replace what the gradient carries and keep the
number of candidates, which shows that the gain comes from the gradient's
information.}
\label{supp-tab:signals}
\end{table}

\section{Detailed Experimental Results}
\label{supp-sec:results}

Every experiment below is tied to a released machine-readable result and a
hash-checked record. We report each experiment against its prespecified threat
model, controls and statistical procedure, then compare the conclusions across
observation grants, corpora, attack paths and published attacks.

\subsection{What the attacker observes}
\label{supp-sec:results-observe}

The main experiment fixes its plan before running. It uses 256 held-out
documents per run and ten client models trained with different random
seeds, each attacked twice with different seeds for the attacks' own random
choices. The primary statistical test fixed in advance is the \emph{two-point
gain test}: at depth six with the public copy of the layers, the joint
attacker's token recovery minus the activation-only attacker's must have
the lower end of its 95\% interval above two percentage points. The
interval comes from a bootstrap that resamples trained models and documents
independently. The protocol also prespecified 47 secondary comparisons, each
tested with a Bonferroni-corrected bootstrap interval. The client's model is
GPT-2 \cite{radford2019gpt2} (124M) on pinned 32-token WikiText-2 passages
\cite{merity2016wikitext}, split after the third or the sixth layer, with 32
public calibration, 128 training and 256 evaluation documents. Training uses
AdamW \cite{loshchilov2019adamw}. The attacker holds either the public first
layers or the client's trained first layers. It attacks the fixed snapshot
described in Section~\ref{supp-sec:method}: in training mode it reads the
activation and the gradient, in inference mode the activation alone. In all, 40 attack configurations (28
without a defence, 12 with one) run against ten trained models at both
depths with both copies of the layers, for 3,200 measurements; the six
activation-only configurations are reused for inference, adding 480, for
3,680 in total, each on the same 256 documents. Sixteen slices ran on two
machines with hash-chained logs; the sharding, the merge and the counts are
in Section~\ref{supp-app:release}.

The gain test passed (Table~\ref{supp-tab:v3primary}). At depth six with the
public layers, the joint attacker recovered 97.38\% of tokens against
94.20\% for the activation-only attacker, 3.17 points more (95\% interval
$[+2.72, +3.64]$). The 47 secondary comparisons cover the joint
attacker against the activation-only attacker, the shuffled-gradient
control, the random-candidate control and the backward-only heuristic, at
both depths, with both copies of the layers and with candidate lists of 8,
16 and 32 tokens, leaving out the gain test itself. In all 47 the joint
attacker came out ahead, with intervals above zero after a Bonferroni
correction for the number of comparisons. The backward-only heuristic,
which reads the gradient alone, recovers 49.98\% of tokens in the same
setting. In the two controls, the shuffled gradient recovers 93.91\% and
the random candidates 93.89\%, both about 0.3 points below the 94.20\%
activation-only result, while replacing the true gradient with a perturbed
one leaves recovery near the true-gradient result, at 97.31\% when we add Gaussian noise
whose standard deviation is 0.1 times the root mean square of each
document's gradient, and 97.33\% when we quantise the gradient to eight
bits. Together, these controls isolate the information contributed by the
returned gradient.

\begin{table}[H]
\centering\small
\begin{tabular}{@{}llcccc@{}}
\toprule
 & & \multicolumn{2}{c}{\textbf{Token accuracy}} &
 \multicolumn{2}{c}{\textbf{Whole-document recovery}} \\
\cmidrule(lr){3-4} \cmidrule(lr){5-6}
\textbf{Depth} & \textbf{Layers held} & \textbf{Activation-only} & \textbf{Joint} &
\textbf{Activation-only} & \textbf{Joint} \\
\midrule
$L{=}3$ & public & 97.39\% & 98.29\% & 43.13\% & 55.51\% \\
$L{=}3$ & trained  & 97.81\% & 98.72\% & 51.41\% & 66.13\% \\
$L{=}6$ & public & \keyresult{94.20\%} & \keyresult{97.38\%} & \keyresult{13.71\%} & \keyresult{37.77\%} \\
$L{=}6$ & trained  & 94.81\% & 98.27\% & 20.23\% & 58.55\% \\
\bottomrule
\end{tabular}
\caption{\textbf{The gradient adds little per token but much per document.}
Token accuracy and whole-document recovery for the two attackers at each
depth and copy of the layers. Each entry pools
5{,}120 document-level scores across the 256 held-out documents.
The same experiment supplies the companion paper's \cite{politis2026layercount}
table; both draw on this one evidence base.}
\label{supp-tab:v3primary}
\end{table}

We also ran paired comparisons against two defences, secret
mixup and replica median. Each defended model is trained separately with its
defence in place.
\emph{Secret mixup} blends each activation with a real activation from an
earlier training batch, three parts true to one part decoy. The client
keeps secret which earlier activation it mixed in, and it corrects the
returned gradient for the decoy. Under it, token recovery ranged from
83.01\% to 90.67\% and whole-document recovery reached at most 1.48\%, while
separate utility measurements showed the geometric-mean perplexity rise by
31.8\% at depth three and 26.9\% at depth six. We also scored annotated
\emph{spans}, the stretches of text that hold a name, a number or an
identifier. Under secret mixup the attacker recovered 47.85--66.15\% of
name spans (716 spans in 249 documents), 93.94--96.50\% of number spans
(297 spans in 142 documents) and 90--100\% of identifier spans (two spans
in two documents), against 67.36--97.44\% of name spans without the
defence. The annotations were fixed before the run, with names marked with
AI assistance. These semantic annotations are not human gold labels.

\emph{Replica median} runs three copies of the server's layers and takes
the median of their outputs, number by number. It protects the integrity
of the server's result and sends the activation unchanged, so token
recovery stayed within two-thirds of a point of the undefended attack.
Whole-document recovery still moved by up to 7.7 points in the most
affected setting (public copy, depth three, SipIt at 128 steps, from
43.77\% to 36.11\%). The defended model is a separate training run, with
three slightly perturbed server copies, so its client layers end slightly
different, and because one wrong token spoils a document, small changes
per token move many whole documents.

The prespecified contrast and all secondary intervals are recorded in
\href{\datasetfile/v3-campaign/review/contrasts.json}{\nolinkurl{v3-campaign/review/contrasts.json}};
per-arm token, document and span results are in
\href{\datasetfile/v3-campaign/review/descriptive_summary.json}{\nolinkurl{v3-campaign/review/descriptive_summary.json}},
and the merged-run integrity check is in
\href{\datasetfile/v3-campaign/merge-verification.json}{\nolinkurl{v3-campaign/merge-verification.json}}.

\subsection{Which text the attacker rebuilds}
\label{supp-sec:results-text}

The same attacker recovers very different amounts on different corpora
(Table~\ref{supp-tab:domain}). We ran the undefended inversion attack on three
pinned corpora, each scored against its own trivial guess: this repository's
own Python source code, synthetic clinical-style notes written from
templates, and a WikiText-2 reference. The trivial guess itself differs: it
scores 2.64\% on Wikipedia and 20.31\% on code, almost eight times higher.
Recovery tracks how regular the text is. The attacker recovers every token
of the templated clinical notes, 64.94\% of the code and 63.25\% of the
Wikipedia text, so corpus choice alone moves recovery by up to 37 points.
Against the corpus-specific guess rates, excess recovery is $+46.78$ points
for code, $+93.75$ points for the clinical notes and $+62.79$ points for
Wikipedia.

Finite-sample floors are 2.3--3.9 points for the 768-position training rows
(Table~\ref{supp-tab:training-floor}) and 0.84--1.90 points for the 3,072-position
inference rows. Every undefended result exceeds its corpus-specific guess by
at least 46 points, so every undefended condition fails the privacy gate.
Added Gaussian noise ($\sigma{=}0.1$) changed mean recovery by at most 1.4
points; the paired intervals include zero for code and Wikipedia and are
exactly zero for the templated clinical corpus.

\begin{table}[H]
\centering\small
\begin{tabular}{@{}lcccc@{}}
\toprule
Corpus & Trivial guess & Recovery & Excess & Verdict \\
\midrule
Source code (own repository) & 20.31\% & \keyresult{64.94\%} & $+46.78$~\pp & \failresult{FAIL} \\
Synthetic clinical-style & 6.25\% & \keyresult{100.00\%} & $+93.75$~\pp & \failresult{FAIL} \\
WikiText-2 reference & 2.64\% & \keyresult{63.25\%} & $+62.79$~\pp & \failresult{FAIL} \\
\bottomrule
\end{tabular}
\caption{\textbf{Corpus choice moves recovery by up to 37 points.}
Undefended inversion (GPT-2, depth six, inference rows, $n{=}3{,}072$).
Each corpus is pinned and scored against its own trivial guess; the floor
for these rows is 0.84--1.90 points. Training rows ($n{=}768$) show the same
direction; their guess rates and floors are in
Table~\ref{supp-tab:training-floor}. Both row counts pool three seeds.}
\label{supp-tab:domain}
\end{table}

\begin{table}[H]
\centering\small
\begin{tabular}{@{}lrrr@{}}
\toprule
Corpus & Guess count & Guess rate & Floor (\pp) \\
\midrule
Source code & 154/768 & 20.05\% & 3.91 \\
Synthetic clinical-style & 48/768 & 6.25\% & 2.60 \\
WikiText-2 reference & 36/768 & 4.69\% & 2.34 \\
\bottomrule
\end{tabular}
\caption{\textbf{Guess rates and floors for the training rows of the domain
battery.} Each row pools 768 token positions across three runs. The floor
is the excess an attack scores at the guess rate; full-precision constants
are in Section~\ref{supp-app:precision}.}
\label{supp-tab:training-floor}
\end{table}

The machine-readable domain results are
\href{\datasetfile/domain-shift-v1/domain-shift-v1-summary.json}{\nolinkurl{domain-shift-v1/domain-shift-v1-summary.json}}
and
\href{\datasetfile/domain-shift-v1/domain-shift-v1-defended-summary.json}{\nolinkurl{domain-shift-v1/domain-shift-v1-defended-summary.json}}.

The evaluator also chooses how long the client trains. A
long-horizon battery ran six defences at 20 and 100 training steps, over
three training seeds, on the \emph{pilot corpus}: 96 pinned WikiText-2
passages from the dataset's training split, which we used while building
the framework and which are separate from the main experiment's held-out
documents (Table~\ref{supp-tab:horizon}). The defences are lightweight stand-ins
for defences in the VFLAIR-LLM collection \cite{vflair}, each a plain
transform. \emph{Added noise} adds Gaussian noise with standard deviation
15 to every number of the activation. \emph{Sparsify-10\%} keeps only the
largest tenth of each activation's numbers. \emph{Mid-lite}, named after
the mutual-information defence but learning nothing, projects the
activation through a fixed random map down to 64 numbers. \emph{Santext},
named after the SanText text-sanitisation defence, swaps each input token,
with probability 0.3, for one of its nearest tokens in the embedding table.
\emph{At-lite}, named after adversarial training but performing none,
shifts every activation by a fixed amount along the direction in which
activations vary most. The corpus's majority token does not occur in the
evaluated passages, giving the battery a fixed 0\% guess rate.

No verdict changed between 20 and 100 steps. Against the declared inversion
attacker, which does not know the projection, mid-lite remained at 0\%
recovery. The 500-step extension shows that mid-lite achieves this result by
largely preventing learning: mean training cross-entropy is 7.42--7.75 versus
4.39--4.44 without a defence. Added-noise recovery rose from 0.3\% in steps
0--7 to 4.2\% in steps 50--99, and sparsify-10\% rose from 17.5\% to 25.9\%.
A document-clustered re-analysis changed no verdict, although the 20-step
added-noise interval ($[+0.47,+1.35]$ points) crossed the one-point threshold.

\begin{table}[H]
\centering\small
\begin{tabular}{@{}lcc@{}}
\toprule
Defence & 20-step excess (\pp) & 100-step excess (\pp) \\
\midrule
none & $+63.02$ & $+58.89$ \\
added noise ($\sigma{=}15$) & $+2.42$ & $+3.69$ \\
sparsify-10\% & $+23.52$ & $+25.03$ \\
\textbf{mid-lite $k{=}64$} & \passresult{$+1.00$} & \passresult{$+0.20$ (PASS, at floor)} \\
santext-30\% & $+46.15$ & $+42.67$ \\
at-lite $\varepsilon{=}15$ & $+63.17$ & $+59.00$ \\
\bottomrule
\end{tabular}
\caption{\textbf{No verdict moves between 20 and 100 training steps.}
Long-horizon defence battery on GPT-2 at depth six (inversion attack,
pilot corpus, mean per-run excess over three training seeds, this
battery's own trivial guess). Mid-lite passes at the floor at both
horizons; every other defence fails.}
\label{supp-tab:horizon}
\end{table}

A 500-step extension ran six trajectories at depth six, with no defence and
with mid-lite, over three training seeds, and two machines reproduced the
saved-prediction analysis and verified every log. The inversion attack
recovered 81.22--81.73\% of tokens with no defence and 0.04--0.85\% with
mid-lite, below its 4.83--4.96\% majority baseline. This extends the same
declared setting to 500 steps and confirms the verdict across three seeds and
two independently reproduced analyses.
The horizon, clustered and 500-step records are, respectively,
\href{\datasetfile/long-horizon-v1/report.json}{\nolinkurl{long-horizon-v1/report.json}},
\href{\datasetfile/clustered-gate-v1/clustered-gate-v1-summary.json}{\nolinkurl{clustered-gate-v1/clustered-gate-v1-summary.json}}
and
\href{\datasetfile/defence-training-500-step-extension/REPRODUCTION.json}{\nolinkurl{defence-training-500-step-extension/REPRODUCTION.json}}.

\subsection{What the validated white-box attack recovers}
\label{supp-sec:results-correct}

The white-box analysis optimises the complete differentiable path from token
embeddings through the client's layers and the true compressor to the observed
message. Twenty PyTorch checks validate the gradient path, work accounting and
capture matching. We evaluated 198 stable-compressor cells across all eleven
internal GPT-2 depths and a separate 99-cell co-trained tier on held-out
final-state inference captures. Each defence compresses the activation.
\emph{PCA compression} sends only the activation's top 64, 128 or 256
principal components out of its 768 numbers (the \emph{width}), fitted once
on calibration text. A \emph{trained bottleneck} starts from the same
projection and adds a small learned correction, trained to defeat probes
that try to read the tokens. Across three training seeds, both methods and
three widths, the attack recovers between 18.85\% and 68.59\% of tokens on
average at every depth (Table~\ref{supp-tab:whitebox}). Each mean averages the
three widths.

\begin{table}[H]
\centering\small
\begin{tabular}{rrrr}
\toprule
Depth & PCA (fixed) & Bottleneck (frozen) & Bottleneck (co-trained) \\
\midrule
1 & 59.16\% & 60.36\% & 61.14\% \\
2 & 66.86\% & \keyresult{68.59\%} & 67.68\% \\
3 & 58.32\% & 58.00\% & 60.24\% \\
4 & 56.88\% & 57.35\% & 60.22\% \\
5 & 55.90\% & 57.02\% & 56.80\% \\
6 & 50.89\% & 52.16\% & 53.09\% \\
7 & 48.94\% & 49.73\% & 48.01\% \\
8 & 41.66\% & 42.05\% & 41.10\% \\
9 & 38.82\% & 37.96\% & 38.24\% \\
10 & 31.01\% & 31.80\% & 30.31\% \\
11 & \keyresult{18.85\%} & 19.67\% & 23.52\% \\
\bottomrule
\end{tabular}
\caption{\textbf{The validated white-box attack recovers 19--69\% of tokens
at every internal depth.} Token recovery (\%), mean over three training
seeds and three widths at each depth. In the first two columns the
compressor stays fixed while the client trains (the bottleneck is trained
beforehand and then frozen), and the attacker reads training-time traffic
with the public copy of the client layers and the true compressor. In the
last column the bottleneck trains alongside the client, and the attacker
reads inference traffic from the final model, holding the client's trained
layers and the final compressor. The stable and co-trained columns report their
separate, fully specified protocols.}
\label{supp-tab:whitebox}
\end{table}

A separate 99-run tier trains the bottleneck alongside the client, then
attacks held-out inference captures from the final state, giving the
attacker the client's trained layers and the trained compressor; every
capture replays exactly from the saved parts. Like the rest of the sweep,
it uses the pilot corpus. We report the stable and co-trained tiers separately,
with their verified cells, means and work counts in
\href{\datasetfile/all-boundary-whitebox-reconstruction/full/summary.json}{\nolinkurl{all-boundary-whitebox-reconstruction/full/summary.json}};
full compute counts and wall times are in Section~\ref{supp-app:precision}.

We also ran the inversion attack on longer texts and on dialogue
(Table~\ref{supp-tab:followups}), at depth six over three training seeds with
no defence or added noise, on code and Wikipedia text at 64 and 128 tokens
and on 32-token dialogue prefixes from 96 dialogues. Undefended recovery
ranges from 76.56\% to 90.22\%; added noise lowers it to 9.96--55.05\%.
Each condition uses three training seeds, and the released records include
its majority-guess rate. Noise-aware whitening lowers inference recovery from
21.65\% to 13.09\% and raises training recovery from 11.46\% to 13.67\%, so
the plain and whitening variants expose different signals.

\begin{table}[H]
\centering\small
\begin{tabular}{lr rrrr}
\toprule
 & & \multicolumn{2}{c}{Training} & \multicolumn{2}{c}{Inference} \\
Corpus & Tokens & None & Added noise & None & Added noise \\
\midrule
Code & 64 & 78.32\% & 21.55\% & 79.70\% & 26.27\% \\
Code & 128 & 87.37\% & 23.11\% & 86.71\% & 32.92\% \\
Wikipedia reference & 64 & 81.77\% & 9.96\% & 82.50\% & 20.82\% \\
Wikipedia reference & 128 & 88.96\% & 14.55\% & 90.22\% & 26.17\% \\
Dialogue & 32 & 76.56\% & 36.72\% & 76.99\% & 55.05\% \\
\bottomrule
\end{tabular}
\caption{\textbf{Longer texts and dialogue leak without a defence; added
noise lowers recovery but does not stop it.} Inversion-attack recovery (\%),
means over three training seeds. Added noise is $\sigma{=}15$. Training and
inference read different signals.}
\label{supp-tab:followups}
\end{table}

The verified raw records and their manifests are packaged under
\href{\datasetfile/phase2-raw-evidence/README.md}{the Phase 2 evidence README}.

\subsection{Which published attack the evaluator runs}
\label{supp-sec:results-published}

We first checked that our port of BiSR matches the reference. Using the same
saved models, inverters, observations and budgets as an archived copy of the
upstream source, and running our port and the reference in separate
processes on 256 fresh 32-token documents at depths three and six, with
inverters trained under three seeds, both produced identical predicted tokens in all 3,072 paired runs;
every seed-level difference was zero. These documents, and the 112
auxiliary articles on which the inverters were trained, come from the pinned
WikiText-103 training split, with the evaluation documents held out from
it; none comes from the official test split, and the selection excludes the
inputs of earlier runs and the WikiText-2 splits. This establishes conformance
for the single-seam setting used by our harness. BiSR's published
\emph{two-cut setting} is a distinct protocol: it cuts GPT-2-large into three
parts and exposes activations at one cut and gradients at the other, whereas
both signals cross the one cut in our harness.

We then compared the port with our joint attack under a plan fixed before
running (Table~\ref{supp-tab:bisr-adapted}). It kept the main experiment's 256
documents, ten trained models, both depths and both copies of the layers,
reused the earlier results as controls, and trained fresh inverters on the
public auxiliary text. Each attack used its own fixed protocol budget. At depth six
with the public layers, the port recovered 93.21\% of tokens against 97.38\%
for the joint attack, 4.17 points fewer (95\% interval $[-4.79, -3.53]$),
with the port lower in every trained model, and the whole interval lies
more than two points below zero. Across fifteen prespecified secondary
contrasts, ten intervals lie below zero, one above and four cross zero.
Notably, the refined port matches the joint attack at depth six with trained
layers ($-0.02$ points, interval $[-0.63,+0.58]$), so no attack dominates
every setting.

\begin{table}[H]
\centering\small
\begin{tabular}{@{}llrrrrr@{}}
\toprule
Depth & Layers held & Activation-only & SipIt-128 & Joint & Inverter & Refined \\
\midrule
3 & Public & 97.39 & 97.45 & 98.29 & 91.90 & 96.16 \\
3 & Trained & 97.81 & 98.89 & 98.72 & 91.90 & 97.79 \\
6 & Public & 94.20 & 95.25 & \keyresult{97.38} & 91.67 & \keyresult{93.21} \\
6 & Trained & 94.81 & 98.14 & 98.27 & 91.67 & 98.25 \\
\bottomrule
\end{tabular}
\caption{\textbf{No attack wins everywhere.} Token accuracy (\%) for the
adapted comparison. Each row averages two attack seeds, ten trained models
and the same 256 documents. Activation-only, SipIt-128 and joint are the
reused controls; the
inverter (SIP) and its refinement (bf) are the adapted attack. The paired
intervals and attack-specific budgets are in the text and supplement.}
\label{supp-tab:bisr-adapted}
\end{table}

The exact conformance verdict is
\href{\datasetfile/bisr-conformance-validation/RESULT.json}{\nolinkurl{bisr-conformance-validation/RESULT.json}};
the adapted comparison and its prespecified intervals are in
\href{\datasetfile/bisr-adapted-comparison/analysis/result.json}{\nolinkurl{bisr-adapted-comparison/analysis/result.json}}
and
\href{\datasetfile/bisr-adapted-comparison/analysis/contrasts.json}{\nolinkurl{bisr-adapted-comparison/analysis/contrasts.json}}.

\subsection{What strength is established}
\label{supp-sec:results-strength}

Attack strength depends on the observation and knowledge grants, corpus,
budget, target and control. Table~\ref{supp-tab:strength-evidence} separates the
shared confirmatory GPT-2 campaign from independent audits and qualification
studies. Together they establish structural leakage, corpus sensitivity,
implementation fidelity, published-attack conformance and module behaviour.

\begin{landscape}
\scriptsize
\setlength{\tabcolsep}{3pt}
\renewcommand{\arraystretch}{1.08}
\begin{longtable}{@{}>{\raggedright\arraybackslash}p{2.2cm}>{\raggedright\arraybackslash}p{2.1cm}>{\raggedright\arraybackslash}p{2.4cm}>{\raggedright\arraybackslash}p{1.7cm}>{\raggedright\arraybackslash}p{1.8cm}>{\raggedright\arraybackslash}p{2.2cm}>{\raggedright\arraybackslash}p{2.6cm}>{\raggedright\arraybackslash}p{2.3cm}>{\raggedright\arraybackslash}p{2.0cm}@{}}
\caption{\textbf{What attack strength and module qualification are established.}
Rows retain their own threat model, corpus, budget and comparator. The first
row is the paper's confirmatory reconstruction result; the remaining rows are
independent audits, controlled real-model studies, implementation conformance
and multi-model qualification experiments.}
\label{supp-tab:strength-evidence}\\
\toprule
Threat model & Module or arm & Positive / negative control & Model & Corpus & Budget & Observed recovery or outcome & Comparator & Evidence tier \\
\midrule
\endfirsthead
\caption[]{What attack strength and module qualification are established (continued).}\\
\toprule
Threat model & Module or arm & Positive / negative control & Model & Corpus & Budget & Observed recovery or outcome & Comparator & Evidence tier \\
\midrule
\endhead
\midrule
\multicolumn{9}{r@{}}{\footnotesize continued overleaf}\\
\endfoot
\bottomrule
\endlastfoot
Passive observer; public or trained client layers & Joint reconstruction, v3 arm & Shuffled-gradient, random-candidate and backward-only controls & GPT-2 124M & 256 held-out 32-token WikiText-2 documents & 10 victim seeds $\times$ 2 attack seeds; $K{=}32$; beam 2 & \keyresult{97.38\%} tokens and \keyresult{37.77\%} exact documents with public layers; 58.55\% exact with trained layers & Activation-only: 94.20\% tokens, 13.71\% exact documents & Confirmatory; shared Paper A/B evidence \\
\addlinespace
Compromised cloud observes forward and returned-gradient channels & \attack{latent-probe} plus exact gradient-support read & Injected-leak calibration; shuffled-label negative control & Qwen3-0.6B & Fixed-width WikiText-2 blocks & 9 independent seeds; 4{,}096 frames per main run & Real/decoy partition exact on \keyresult{4{,}096/4{,}096} frames; frequent-token effect about $+0.92$~\pp & Forward-only gate passed; shuffled controls at floor & Independent real-system audit \cite{politis2026privacyfailure} \\
\addlinespace
Passive seam observer; public client layers & \attack{seam-inversion} & Corpus-specific majority guess; undefended positive target & GPT-2 124M & Source code, templated clinical notes, WikiText-2 & 3 seeds; $n{=}3{,}072$ inference positions per corpus & \keyresult{64.94\%}, \keyresult{100.00\%}, \keyresult{63.25\%} & Guess rates 20.31\%, 6.25\%, 2.64\% & Controlled real-model qualification \\
\addlinespace
White-box observer knows the compressor & \attack{seam-inversion} white-box path & Twenty gradient-path and capture-matching checks & GPT-2 124M & Pinned pilot WikiText-2 & 198 stable and 99 co-trained runs; 128 optimisation steps & \keyresult{18.85--68.59\%} mean token recovery over every internal depth & Stable and co-trained tiers use fully specified protocols & Controlled real-model qualification \\
\addlinespace
Single-seam observer with BiSR grants & \attack{bisr-baseline} & Archived reference implementation on identical fresh inputs & GPT-2 124M & Fresh WikiText-103 for conformance; main held-out set for comparison & 3{,}072 paired predictions; 20 inverter epochs; 32 backward + 800 forward refinement steps & Reference agreement on \keyresult{3{,}072/3{,}072}; adapted attack 93.21\% at depth six/public & Joint arm 97.38\%; attack-specific fixed budgets & Reproduced conformance + adapted comparison \\
\addlinespace
Passive seam observer across training horizons & \attack{seam-inversion} & Undefended positive target; majority control; mid-lite protected contrast & GPT-2 124M & Pinned pilot WikiText-2 & 20/100 steps, 3 seeds & Undefended excess $+63.02/+58.89$~\pp; mid-lite $+1.00/+0.20$~\pp & Same attacker on undefended and protected captures & Controlled horizon qualification \\
\addlinespace
Observer of an obfuscated inference wire & \attack{wire-eval} & Unobfuscated baseline and shared-key identity-restoration arm & Mistral-7B-Instruct & Pinned E9 session inputs & 3 paired deployment sessions & Baseline 50\% top-1; obfuscated wire \keyresult{0\%}; held-out loss unchanged within precision & Shared-key arm restores identity & Multi-model defence qualification \\
\addlinespace
Passive activation observer & SipIt qualification arm & Undefended target; bf16/fp32 precision twins & Mistral-7B-Instruct & 10 victim documents per depth and precision & Depths 1/4/8; bf16/fp32; effective $n{=}10$ per cell & \keyresult{77.03--84.53\%} mean token recovery & Precision twins differ by at most 1.41 points & Multi-model attack qualification \\
\addlinespace
Known-pair, stale-key and sharding attackers & \attack{known-prefix}, \attack{stale-key}, \attack{sharded} & Label-free band; fresh-key solve; static and staggered arms & Qwen3-0.6B, Qwen3.5-27B, Mistral-NeMo-12B & Pinned E8/E-R inputs & Measured $K$ grids; repeated solve seeds and sessions & Known-prefix $K_{90}\!\approx\!232$--240; stale key 0.94--2.45\% vs fresh 61.20--65.31\%; sharding 54.95--67.86\% & Label-free band and fresh/static controls & Multi-model attack qualification \\
\end{longtable}
\end{landscape}

Table~\ref{supp-tab:strength-evidence} distinguishes controlled recovery
measurements from multi-model qualification and states the threat model,
budget, control and measured outcome for every protocol.

\paragraph{Cumulative evidence.}
Taken together, the programme supports the following statement:
LLM-Attack Suite has exposed consequential leakage under several independent
protocols: exact structural leakage in a deployed Qwen split-training system,
high text reconstruction in a controlled GPT-2 campaign, and additional
corpus-, placement-, depth- and defence-dependent failures. Its broader Qwen
and Mistral campaigns provide qualification evidence that different modules
have been exercised against multiple models and defences, including expected,
negative and privacy-breaking outcomes.

Table~\ref{supp-tab:module-validation} accounts for all 64 modules and
records the evidence tier, controls and status of each.

\subsection{Live agentic qualification}
\label{supp-sec:results-agentic}

A live tier used Qwen3-0.6B \cite{yang2025qwen3} as the local planner under
the same sandbox, permission checks, side-effect logging and budgets as the
synthetic tier. With reasoning mode disabled, all ten prespecified settings
produced measurable actions. No policy violation was recorded in the twenty
fixed scenario trials. All three sanctioned peer-request settings were
blocked in both trials, giving six false blocks; the result therefore
establishes successful planner/tool integration and identifies systematic
over-blocking of sanctioned requests. The full measurements and planner traces
are in
\href{\datasetfile/agentic-nonthinking-decoding/raw/FOLLOWUP.json}{\nolinkurl{agentic-nonthinking-decoding/raw/FOLLOWUP.json}}
and
\href{\datasetfile/agentic-nonthinking-decoding/raw/planner-traces.jsonl}{\nolinkurl{agentic-nonthinking-decoding/raw/planner-traces.jsonl}}.

\subsection{Framework operation across protocols}
\label{supp-sec:results-framework}

The completed campaigns exercise the workflow in
Figure~\ref{fig:framework} at several stages. The main campaign reused
captures across attack configurations and assembled 3,680 measurements,
including 480 explicitly reused inference measurements, from sixteen
shards. Its
\href{\datasetfile/v3-campaign/merge-verification.json}{merge-verification record}
and
\href{\datasetfile/v3-campaign/EVIDENCE-RELEASE.md}{release and re-verification guide}
document the assembly and provenance checks. The BiSR conformance study
tests implementation fidelity against a reference executable; its adapted
comparison demonstrates how another attack can join a frozen experiment.
The white-box study exercises defence-aware reconstruction across internal
split depths, and the live agentic study connects a local planner to the
sandbox's action and policy records.

The Qwen and Mistral campaigns extend module qualification across the models
and defences recorded in Table~\ref{supp-tab:module-validation}. Together, these
studies establish reuse of the framework across reconstruction, robustness,
obfuscation and agentic protocols. Their shared methodological role is to
connect an explicit attacker contract to controls, measured outcomes and
retained evidence. This is the basis for adding a new attack or target to
the programme while preserving the interpretation of earlier results.

\subsection{Validation of the evaluation methodology}
\label{supp-sec:results-method-validation}

We tested whether the evaluation code preserves the intended meaning of
recovery measurements and rejects specified errors in their supporting
evidence. The design combines known-answer fixtures, independent numerical
calculations and fault injection with metamorphic tests: checks of prescribed
relationships between outputs under controlled changes to the inputs
\cite{xie2011metamorphic}. For example, changing text available only to the
scorer must leave the attack's predictions unchanged, while removing a
required positive control must prevent a candidate pass.

We fixed a selection of 25 existing development-test functions before
re-executing them from the reviewed framework revision on CPU, retaining
their original assertions and numerical tolerances. Parameterised inputs
expanded the selection to 37 cases across seven measurement properties
(Table~\ref{supp-tab:measurement-validation}). The inputs were constructed
token sequences, numerical arrays, small randomly initialised models and
synthetic evidence records.

\begin{table}[H]
\centering\small
\renewcommand{\arraystretch}{1.12}
\begin{tabular}{@{}>{\raggedright\arraybackslash}p{0.18\textwidth}>{\raggedright\arraybackslash}p{0.31\textwidth}>{\raggedright\arraybackslash}p{0.36\textwidth}>{\raggedleft\arraybackslash}p{0.07\textwidth}@{}}
\toprule
Property & Controlled intervention & Expected response & Cases \\
\midrule
MV1: recovery metrics & Known token errors, absent spans and misplaced matching text & Exact token and span counts; missing spans stay missing; misplaced matches score zero & \keyresult{3/3} \\
\addlinespace
MV2: score conformance & Independent Wilson calculations and threshold crossings & Numerical agreement within tolerance; prescribed boundary verdicts & \keyresult{4/4} \\
\addlinespace
MV3: control eligibility & Retain, remove, weaken or make a linked control unreadable & Backed candidate pass retained; other candidates marked insufficient evidence & \keyresult{4/4} \\
\addlinespace
MV4: observation boundary & Change scorer-only truth; supply forbidden truth or malformed observations & Byte-identical predictions after truth change; invalid inputs rejected & \keyresult{2/2} \\
\addlinespace
MV5: solver fidelity & Independent GRU and AdamW calculations; white-box head checks & Outputs and derivatives agree; correct layer prefixes and nonzero embedding gradients & \keyresult{3/3} \\
\addlinespace
MV6: resampling & Constant data, correlated document duplication and invalid matrices & Known intervals retained in the fixed cases; invalid matrices rejected & \keyresult{6/6} \\
\addlinespace
MV7: evidence validation & Two intact cases, ten invalid report states and three retained-hash corruptions & Intact evidence accepted; all specified faults rejected & \keyresult{15/15} \\
\bottomrule
\end{tabular}
\caption{\textbf{All 37 selected validation cases matched their specified outcomes.}
The last column counts matched outcomes over executed cases, including
parameterised inputs; rejecting an
invalid input is a successful check. MV1--MV7 identify the case groups in the
measurement-validation supplement described in Section~\ref{supp-app:release}.}
\label{supp-tab:measurement-validation}
\end{table}

There were no failures or skips. In the dependence stress case, increasing
perfectly correlated document columns
from two to 500 preserved the interval $[0,1]$ rather than treating the
copies as independent evidence. The BiSR prediction process produced
byte-identical outputs after its scorer-only truth file changed and rejected
a manifest that supplied that truth to the predictor. These results establish
conformance of the selected measurement procedures on their fixed cases.
They complement the external-reference comparison in
Section~\ref{supp-sec:results-published} and the real-model leakage measurements.

\section{Release and Reproduction}
\label{supp-app:release}

The main experiment ran as sixteen deterministic slices across two machines,
merged with hash verification, so that depth is never tied to one host;
the merge checks that every prediction record and every log verifies and
that the runs together equal the planned set. Its 860 binary-tier
references are hash-verified against the pinned baseline release, with no
missing or mismatched files. The Hugging Face dataset at
\url{\dataseturl} holds the predictions, logs and captures of the main
experiment, the domain, horizon and clustered analyses, the two distinct
agentic packages, the positive controls, and the code, protocols and inputs.
Tag \texttt{v2.0} collects the retained campaign evidence, the revised manuscript
and the measurement-validation records identified below. It reuses the
unchanged scientific evidence from v1.0 and changes branding and paper
packaging only, with no new experiments or runtime changes. Experimental files
retain their original hashes; source snapshots preserve the code needed for
reproduction independently of the former Git history.

The manuscript, reviewed 64-module taxonomy and released code snapshot
have distinct roles; Table~\ref{supp-tab:identifiers} records each
identifier and scope. The GitHub remote
\url{https://github.com/setloop-io/LLM-Attack-Suite/tree/v2.0} is the canonical
source release, and the tagged Hugging Face release is the reproducibility artifact.
The released command-line tool offers self-test, module listing, scorecard,
coverage and profile reports, and the synthetic tier needs no deep-learning
stack. Tables~\ref{supp-tab:release} and \ref{supp-tab:identifiers} give the principal
release locations and identifiers.

\begin{table}[H]
\centering\footnotesize
\renewcommand{\arraystretch}{1.15}
\begin{tabular}{@{}>{\raggedright\arraybackslash}p{3.5cm}>{\raggedright\arraybackslash}p{10.5cm}@{}}
\toprule
Evidence & Path in the tagged dataset release \\
\midrule
Release root and manuscript &
\href{\datasetrelease}{\texttt{v2.0}};
\href{\datasetrelease/paper}{\nolinkurl{paper/}} \\
\addlinespace
Main campaign and controls &
\href{\datasetrelease/v3-campaign}{\nolinkurl{v3-campaign/}};
\href{\datasetrelease/binary-tier}{\nolinkurl{binary-tier/}};
\href{\datasetrelease/positive-controls}{\nolinkurl{positive-controls/}} \\
\addlinespace
Inputs and protocols &
\href{\datasetrelease/corpora}{\nolinkurl{corpora/}};
\href{\datasetrelease/protocols}{\nolinkurl{protocols/}} \\
\addlinespace
Domain and horizon analyses &
\href{\datasetrelease/domain-shift-v1}{\nolinkurl{domain-shift-v1/}};
\href{\datasetrelease/long-horizon-v1}{\nolinkurl{long-horizon-v1/}};
\href{\datasetrelease/clustered-gate-v1}{\nolinkurl{clustered-gate-v1/}} \\
\addlinespace
Validated white-box results &
\href{\datasetrelease/all-boundary-whitebox-reconstruction/original36}{\nolinkurl{all-boundary-whitebox-reconstruction/original36/}};
\href{\datasetrelease/all-boundary-whitebox-reconstruction/full}{\nolinkurl{all-boundary-whitebox-reconstruction/full/}} (31 files);
\href{\datasetrelease/phase2-raw-evidence}{\nolinkurl{phase2-raw-evidence/}} (four files, 7.15\,GB) \\
\addlinespace
BiSR and adapted comparison &
\href{\datasetrelease/bisr-conformance-validation}{\nolinkurl{bisr-conformance-validation/}};
\href{\datasetrelease/bisr-conformance-run-artifacts}{\nolinkurl{bisr-conformance-run-artifacts/}};
\href{\datasetrelease/bisr-adapted-comparison}{\nolinkurl{bisr-adapted-comparison/}} \\
\addlinespace
Training extension &
\href{\datasetrelease/defence-training-500-step-extension}{\nolinkurl{defence-training-500-step-extension/}} \\
\addlinespace
Agentic evidence &
live planner: \href{\datasetrelease/agentic-live-v1}{\nolinkurl{agentic-live-v1/}};
non-thinking amendment: \href{\datasetrelease/agentic-nonthinking-decoding}{\nolinkurl{agentic-nonthinking-decoding/}} \\
\addlinespace
Indexes and checksums &
\href{\datasetfile/INDEX.md}{\nolinkurl{INDEX.md}};
\href{\datasetfile/REPRODUCE.md}{\nolinkurl{REPRODUCE.md}};
\href{\datasetfile/MANIFEST.sha256}{\nolinkurl{MANIFEST.sha256}};
\href{\datasetfile/BINARY-VERIFICATION.json}{\nolinkurl{BINARY-VERIFICATION.json}};
\href{\datasetfile/v3-campaign/EVIDENCE-RELEASE.md}{\nolinkurl{v3-campaign/EVIDENCE-RELEASE.md}} \\
\bottomrule
\end{tabular}
\caption{\textbf{Principal release locations.} Current paths in the cleaned
\texttt{v2.0} tree. Use the tagged paths; former Git commit and release URLs
are not navigation targets for this release.}
\label{supp-tab:release}
\end{table}

\begin{table}[H]
\centering\footnotesize
\renewcommand{\arraystretch}{1.15}
\begin{tabular}{@{}>{\raggedright\arraybackslash}p{3.6cm}>{\raggedright\arraybackslash}p{10.4cm}@{}}
\toprule
Item & Identifier or scope \\
\midrule
Combined manuscript & \nolinkurl{paper/}; the concise paper and full supplement are supplied as one PDF and a self-contained arXiv source ZIP \\
\addlinespace
Reviewed framework taxonomy & \nolinkurl{source-snapshots/reviewed-framework-64.tar.gz}; 64 registered modules; historical source identifier \texttt{6c071a68ae5134522d0a4cd3d7d0fa67acabdfa8} \\
\addlinespace
Dataset code snapshot & \nolinkurl{code/} and \nolinkurl{source-snapshots/base-framework-code.tar.gz}; historical base source identifier \texttt{a3eb53fca5d6395da4c4377e9d8787676c66ae6a}, supplemented by experiment-specific source archives \\
\addlinespace
Maintained framework & GitHub tag \texttt{v2.0} retains the v1.0 runtime and 65 registered modules; the additional \attack{secure-compute-audit} module is outside the paper's reviewed 64-module qualification matrix \\
\addlinespace
Models and corpus & GPT-2 124M \texttt{openai-community/gpt2}, revision \texttt{607a30d783dfa663caf39e06633721c8d4cfcd7e}; Qwen3-0.6B \texttt{Qwen/Qwen3-0.6B}, revision \texttt{c1899de289a04d12100db370d81485cdf75e47ca}; released \nolinkurl{corpora/} and \nolinkurl{protocols/} are the binding text inputs \\
\addlinespace
Seeds & main experiment: ten trained-model seeds and two attack seeds; white-box and horizon: 11, 22, 33; BiSR check: 31415, 27182, 16180; adapted comparison: attack seeds 17, 29; bootstrap: 93281 \\
\addlinespace
Licence metadata & Framework and release packaging: MIT, with \nolinkurl{LICENSE} supplied. Third-party components retain their own licence terms \\
\bottomrule
\end{tabular}
\caption{\textbf{Revision, model and access identifiers.} The distinct
source components are identified separately so that the manuscript, reviewed
taxonomy and released code snapshot can be reproduced without relying on old Git refs.}
\label{supp-tab:identifiers}
\end{table}

\paragraph{Measurement-validation supplement.}
The additional validation run in
Section~\ref{supp-sec:results-method-validation} is supplied in
\href{\datasetrelease/validation/measurement-20261001}{\nolinkurl{validation/measurement-20261001/}}
within release \texttt{v2.0}, unchanged from v1.0. The corresponding path in the maintained
source repository is \nolinkurl{validation/measurement-20261001/}. Its original
selection, expected outcomes and per-case observations remain in
\nolinkurl{selection.json} and \nolinkurl{run/results.json}; the original
\nolinkurl{MANIFEST.sha256} is preserved unchanged. The package also contains
\nolinkurl{source-snapshot.tar.gz}, which matches all 123 retained source-file
hashes, and \nolinkurl{reproduce_from_snapshot.py}, which reproduces the frozen
selection without Git history. Added packaging files have separate digests in
\nolinkurl{SNAPSHOT.sha256}.

\section{Names in the Released Files}
\label{supp-app:labels}

The application is named LLM-Attack Suite. Its earlier name,
LLM-Attacker, may remain in software identifiers, paths and historical
release metadata. These retained identifiers refer to this suite, not
the autonomous-driving framework of Mei et al.\ \cite{mei2025llmattacker}.
The naming change does not alter the frozen experimental evidence.

The released files use internal names for the experiments and attacks. The
main experiment is the
\emph{v3} campaign, and the adapted comparison is the \emph{v4} adaptation.
The joint attack is \texttt{joint\_true\_k32}, the activation-only attack is
\texttt{forward\_k32}, the shuffled-gradient and random-candidate controls
are \texttt{joint\_shuffled\_k32} and \texttt{joint\_random\_k32}, the
backward-only heuristic is \texttt{backward}, and SipIt with 32, 64 and 128
steps per token is \texttt{sipit\_steps32}, \texttt{sipit\_steps64} and
\texttt{sipit\_steps128}. The BiSR implementation
check is the \emph{C-A} route, and the inverter and its refinement are
\emph{SIP} and \emph{bf}. The pass rule's current version is
\texttt{frozen-gate.v2} and the scorecard policy is
\texttt{scorecard-policy.v3}. The benchmark profiles are
\texttt{benchmark.split-privacy.v2},
\texttt{benchmark.split-integrity-availability.v1},
\texttt{benchmark.genai-app.v1} and \texttt{benchmark.agentic.v12}, and the
observation points of Figure~\ref{supp-fig:anatomy} and Table~\ref{supp-tab:surfaces} are S1--S14.

\section{Full Module List and Full-Precision Numbers}
\label{supp-app:framework}
\label{supp-app:precision}

\paragraph{Attack modules and record.}
Each run writes one JSON record with the fields \texttt{schema},
\texttt{config}, \texttt{threat\_model}, \texttt{provenance},
\texttt{results} and \texttt{summary}. The coverage report marks each threat
as COVERED, PARTIAL, UNMEASURED or OUT\_OF\_SCOPE. The benchmark profiles
group the threats: \texttt{benchmark.split-privacy.v2} covers the split
privacy families with tenant isolation, telemetry leakage and the router
channel; \texttt{benchmark.split-integrity-availability.v1} covers protocol
integrity, malicious client and server, poisoning, resource exhaustion, key
abuse and supply chain; \texttt{benchmark.genai-app.v1} covers the OWASP LLM
list; and \texttt{benchmark.agentic.v12} covers the OWASP agentic list and is
out of scope by default. Table~\ref{supp-tab:taxonomy} lists the 64 modules by
family; the count is exact at the reviewed revision
\texttt{submission-candidate-20260922-r2}.

{\small
\begin{longtable}{@{}>{\raggedright\arraybackslash}p{2.9cm}>{\raggedright\arraybackslash}p{5.3cm}>{\raggedright\arraybackslash}p{1.1cm}>{\raggedright\arraybackslash}p{5.3cm}@{}}
\caption{Attack families in the reviewed LLM-Attack Suite taxonomy (64
registered modules at tag \texttt{submission-candidate-20260922-r2}).
Modes: T = training, I = inference. Entries describe implemented
capabilities; Table~\ref{supp-tab:module-validation} records each module's evidence
tier.}
\label{supp-tab:taxonomy}\\
\toprule
Family & Modules & Modes & Capability / evidence scope \\
\midrule
\endfirsthead
\caption[]{Attack families in LLM-Attack Suite (continued).}\\
\toprule
Family & Modules & Modes & Capability / evidence scope \\
\midrule
\endhead
\midrule
\multicolumn{4}{r@{}}{\footnotesize continued overleaf}\\
\endfoot
\bottomrule
\endlastfoot
Trained inversion &
\attack{latent-probe}, \attack{latent-sensitivity},
\attack{vec2text-inversion}, \attack{seam-inversion} &
T, I &
Trained and optimisation-based reconstruction; the main experiment and the
validated white-box results have their own observations and budgets \\
\addlinespace[2pt]
Matching attacks &
\attack{latent-matching} &
T &
Centroid and signature matching; token and order results are kept
separately \\
\addlinespace[2pt]
Key / budget solves &
\attack{accumulation}, \attack{stale-key}, \attack{sharded},
\attack{leak-accumulation}, \attack{subspace-joint}, \attack{wire-eval},
\attack{seed-disclosure} &
T, I &
Known-pair budgets, cross-epoch pooling and key-lifetime controls \\
\addlinespace[2pt]
Known-plaintext &
\attack{known-prefix}, \attack{alignment-search} &
T, I &
Known-prefix solves and realignment under declared label and alignment
assumptions \\
\addlinespace[2pt]
Gradient / output &
\attack{gradient-inversion}, \attack{output-inversion},
\attack{label-scoring}, \attack{batch-label-recon} &
T, I &
Gradient and output reconstruction and gradient-size property scoring \\
\addlinespace[2pt]
Membership &
\attack{membership}, \attack{lira-membership} &
T &
Likelihood-ratio and relative-likelihood scoring with null-world controls
under a local protocol \\
\addlinespace[2pt]
Structure / blind &
\attack{ica-bss} &
T, I &
Higher-order blind-source separation beyond covariance attacks \\
\addlinespace[2pt]
Active / malicious &
\attack{active-cloud}, \attack{malicious-server} &
T, I &
Active-server interventions and steering under declared observation \\
\addlinespace[2pt]
Model-IP theft &
\attack{body-extraction} &
T, I &
Query-driven functional distillation and weight-similarity measurement \\
\addlinespace[2pt]
Hardware / link &
\attack{gpu-memory}, \attack{kv-timing}, \attack{wire-shape} &
T, I &
Allocator-residue and contention probes under a co-resident-tenant threat
model \\
\addlinespace[2pt]
Reasoning theft &
\attack{reasoning-replay}, \attack{trace-inversion} &
I &
Replay of opaque reasoning blobs and hidden-trace synthesis, scored over an
answer-only control \\
\addlinespace[2pt]
Integrity / availability &
\attack{protocol-integrity}, \attack{key-session-abuse},
\attack{resource-exhaustion}, \attack{poisoning-backdoor},
\attack{malicious-client}, \attack{supply-chain}, \attack{replica-integrity} &
T, I &
Protocol mutation, key-lifecycle abuse, poisoning and availability arms
against defended and vulnerable simulators \\
\addlinespace[2pt]
Tenant isolation / composition &
\attack{multi-tenant-isolation}, \attack{artifact-telemetry-leak},
\attack{router-side-channel}, \attack{privacy-composition},
\attack{view-privacy}, \attack{representation-privacy} &
T, I &
Cross-tenant channels, store canary sweeps, router-trace classification and
composition curves \\
\addlinespace[2pt]
Application profile (GenAI) &
\attack{prompt-injection}, \attack{hidden-context-exposure},
\attack{rag-vector-abuse}, \attack{output-sink-injection},
\attack{misinformation-integrity} &
I &
Injection scored by policy-layer effect; retrieval and grounding checks \\
\addlinespace[2pt]
Agentic profile (OWASP ASI) &
\attack{agent-goal-hijack}, \attack{tool-misuse},
\attack{identity-privilege-abuse}, \attack{agentic-supply-chain},
\attack{unexpected-code-execution}, \attack{memory-context-poisoning},
\attack{insecure-agent-communication}, \attack{cascading-failure},
\attack{human-agent-trust}, \attack{rogue-agent}, \attack{cross-agent-deputy} &
I &
Seeded agent sandbox with deny-by-default permissions; attacks scored by
policy-layer effect, each defence checked by single-flag ablation \\
\addlinespace[2pt]
Deployment / eval &
\attack{remote-fetch}, \attack{mitm-proxy}, \attack{max-effort},
\attack{gradient-hardening}, \attack{two-seam-delegation},
\attack{bisr-baseline} &
T, I &
Fleet probing, a relay that forwards traffic unchanged and logs a copy, and
attacker-optimisation distributions \\
\end{longtable}
}

\paragraph{Validation tiers and artifacts.}
Table~\ref{supp-tab:module-validation} accounts for every module at the reviewed
64-module revision and assigns its primary evidence class. \emph{Reproduced}
records agreement with an external reference on fixed inputs; \emph{adapted}
records a prior attack family implemented under a declared local contract;
\emph{qualification-tested} records a framework-native module run against a
real model, capture or deployment; \emph{synthetic} records qualification on a
deterministic fixture or simulator; and \emph{unmeasured/helper} identifies
orchestration components whose outputs are provenance and budget records.
Status \texttt{Q-REAL} denotes real-model or deployment qualification,
\texttt{Q-SYN} synthetic qualification, \texttt{PARTIAL} a mixed evidence
tier, \texttt{UNMEAS} a pending target measurement, and \texttt{HELPER}
orchestration support. Protected and null outcomes qualify a module when they
match their controls. Deployment claims require \texttt{Q-REAL} evidence under
the corresponding deployment protocol.

Artifact keys keep the matrix readable. \texttt{PA/V3} is the shared Paper A/B
\nolinkurl{v3-campaign/}; \texttt{PB/PC}, \texttt{PB/DS}, \texttt{PB/LH},
\texttt{PB/WB}, \texttt{PB/BC}, \texttt{PB/BA} and \texttt{PB/AG} are the
current release's positive controls, domain, horizon, white-box, BiSR
conformance, BiSR adaptation and agentic packages listed in
Table~\ref{supp-tab:release}. \texttt{P1/AUD} and \texttt{P1/E8--ER} refer to the
frozen audit and historical qualification artifacts indexed by the Paper~1
release \cite{politis2026privacyfailure}. \texttt{T/name} is the named
module's deterministic fixture and self-test in the reviewed code tag. The
artifact column names evidence, not merely the source file that registers the
module.

\begin{landscape}
\scriptsize
\setlength{\tabcolsep}{3pt}
\renewcommand{\arraystretch}{1.08}
\begin{longtable}{@{}>{\raggedright\arraybackslash}p{2.8cm}>{\raggedright\arraybackslash}p{1.8cm}>{\raggedright\arraybackslash}p{2.0cm}>{\raggedright\arraybackslash}p{2.1cm}>{\raggedright\arraybackslash}p{2.7cm}>{\raggedright\arraybackslash}p{2.8cm}>{\raggedright\arraybackslash}p{2.4cm}>{\raggedright\arraybackslash}p{1.8cm}>{\raggedright\arraybackslash}p{1.7cm}@{}}
\caption{\textbf{Validation matrix for all 64 reviewed modules.} ``Expected''
is the direction fixed by the module contract; ``observed'' is the retained
qualification outcome, including negative and protected results. Qualification
tests whether the implementation behaves as specified; tier and status
distinguish real-target, synthetic and helper evidence.}
\label{supp-tab:module-validation}\\
\toprule
Module & Tier & Model / target & Defence & Expected outcome & Observed outcome & Positive / negative control & Artifact & Status \\
\midrule
\endfirsthead
\caption[]{Validation matrix for all 64 reviewed modules (continued).}\\
\toprule
Module & Tier & Model / target & Defence & Expected outcome & Observed outcome & Positive / negative control & Artifact & Status \\
\midrule
\endhead
\midrule
\multicolumn{9}{r@{}}{\footnotesize continued overleaf}\\
\endfoot
\bottomrule
\endlastfoot

\multicolumn{9}{@{}l}{\textbf{Reproduced against an external reference (1)}}\\
\midrule
\attack{bisr-baseline} & Reproduced & GPT-2, one seam & None; public / trained layers & Local port and archived reference emit identical predictions; adapted refinement remains separately budgeted & \keyresult{3{,}072/3{,}072} paired predictions identical; adapted depth-6/public recovery 93.21\% & Reference executable on identical inputs; shared joint arm 97.38\% & \texttt{PB/BC}, \texttt{PB/BA} & \texttt{Q-REAL} \\

\midrule
\multicolumn{9}{@{}l}{\textbf{Adaptations of prior attack families (10)}}\\
\midrule
\attack{vec2text-inversion} & Adapted & Generated seam world & Gaussian seam noise & Learned decoder beats nearest-neighbour decoding and degrades with noise & 94.8\% noiseless and 91.1\% at $\sigma{=}2$ in the fixture & Nearest-neighbour and majority controls & \texttt{T/name} & \texttt{Q-SYN} \\
\attack{seam-inversion} & Adapted & GPT-2 124M & None, noise, mixup, PCA / bottleneck & Undefended recovery exceeds the corpus control; defence-aware path can expose residual leakage & 63.25--100\% across domains; validated white-box path 18.85--68.59\% across depths & Corpus majority; undefended target; matched path checks & \texttt{PB/DS}, \texttt{PB/WB}, \texttt{PB/PC} & \texttt{Q-REAL} \\
\attack{gradient-inversion} & Adapted & Qwen3-0.6B & None & Multi-restart DLG++ measures recovery across shallow and deep boundaries & 1.77\%, 0.31\%, 0.21\% at depths 1, 4, 8 over 90 cells & Restarts, depths and random initialisations & \texttt{P1/E8-DLG} & \texttt{Q-REAL} \\
\attack{output-inversion} & Adapted & Qwen3-0.6B; Mistral-7B & None / output-side exposure & Output features should permit recovery above a shuffled or majority control when informative & Attack-specific qualification completed on Qwen3-0.6B and Mistral-7B & Shuffled-output and majority controls & \texttt{P1/E8-OUT} & \texttt{Q-REAL} \\
\attack{label-scoring} & Adapted & Qwen3-0.6B & Split fine-tuning / FedAvg & Gradient-norm score separates token-presence properties above AUC 0.5 & Property AUC 0.979 and 0.984 in the corrected three-seed study & Label shuffle, AUC 0.5 and corrected seed sampler & \texttt{P1/E-A4} & \texttt{Q-REAL} \\
\attack{batch-label-recon} & Adapted & Generated linear body & None & Planted labels are recovered; null system remains at control & Closed-form fixture exceeds 95\% & Planted-label positive and null control & \texttt{T/name} & \texttt{Q-SYN} \\
\attack{lira-membership} & Adapted & GPT-2 124M & Undefended positive target & Likelihood-ratio membership score separates IN and OUT records; null world stays at AUC 0.5 & Offline positive control AUC \keyresult{1.00} on 48 IN / 48 OUT records; low-FPR grid is unresolved at this size & Null-world scores and AUC 0.5 & \texttt{PB/PC} & \texttt{Q-REAL} \\
\attack{body-extraction} & Adapted & Generated teacher / student body & None & Query-driven student improves held-out functional agreement; weight similarity is kept separate & Fixture raises held-out $R^2$, cosine and logit agreement over an untrained student & Held-out queries and random-student control & \texttt{T/name} & \texttt{Q-SYN} \\
\attack{kv-timing} & Adapted & Cache/timing fixture & Cache isolation / cold cache & Warm and cold time-to-first-token distributions separate only when the cache leaks & Warm/cold and speculative-acceptance fixtures separate; hidden-acceptance arm reports INFO & Cold, randomized-prefix and hidden-telemetry controls & \texttt{T/name} & \texttt{Q-SYN} \\
\attack{trace-inversion} & Adapted & Generated reasoning tasks & Answer-only interface & Calibration, search and refinement improve trace consistency over an answer-only or surrogate-style control & Refined/best-of-four traces reach consistency 1.0 in the fixture & Answer-only, surrogate style and uncalibrated arms & \texttt{T/name} & \texttt{Q-SYN} \\

\midrule
\multicolumn{9}{@{}l}{\textbf{Framework-native modules qualified on real models, captures or deployments (15)}}\\
\midrule
\attack{latent-probe} & Qualification & Qwen3-0.6B; 35B-A3B audit tier & Rotated or permuted latents and decoy rows & Calibrated probe detects planted or observable structure and stays at floor after label shuffle & Paper~1 audit exposes the real/decoy partition on every frame; content effect about $+0.92$~\pp & Injected leak, shuffled labels and label-free band & \texttt{P1/AUD}, \texttt{P1/E8} & \texttt{Q-REAL} \\
\attack{latent-sensitivity} & Qualification & Qwen3-0.6B; 35B-A3B audit tier & Chaff, gauge and known-plaintext compromise & Recovery changes with the compromised fraction; zero-compromise arm stays at control & Historical fraction sweeps retained both breaking and non-breaking cells & Zero fraction, random gauge and chaff-retain controls & \texttt{P1/E8} & \texttt{Q-REAL} \\
\attack{latent-matching} & Qualification & Qwen3-0.6B; 35B-A3B audit tier & Rotation, decoys and row reordering & Token and order endpoints exceed their own chance controls only when signatures survive & Real battery retained a negative result; synthetic order-and-decoy excess spans 0.52--5.21 points & Chance order, shuffled labels and constant-token control & \texttt{P1/AUD}, \texttt{T/name} & \texttt{Q-REAL} \\
\attack{accumulation} & Qualification & Qwen/Mistral E-R tiers & Per-epoch rotation; mid-lite & More labelled pairs improve the epoch-local solve; protected contrast remains low & Historical E-R budget sweeps retain both breaking and non-breaking cells; mid-lite remains at 0--0.5\% & Label-free band, epoch-local solve and mid-lite & \texttt{P1/ER} & \texttt{Q-REAL} \\
\attack{stale-key} & Qualification & Qwen3.5-27B & Per-epoch rotation & Stale keys fail on a new epoch; fresh labelled pairs recover the mapping & Stale recovery 0.94--2.45\%; fresh $E{=}1024$ recovery 61.20--65.31\% & Fresh-key solve and label-free band & \texttt{P1/ER4} & \texttt{Q-REAL} \\
\attack{sharded} & Qualification & Qwen3.5-27B; Mistral-NeMo-12B & Static and staggered sharding & A qualified schedule keeps recovery near label-free across blocks and sessions & Static 56.98--67.86\%; staggered 47.58--65.13\%; independent 12B replication 54.95--65.68\% & Label-free band; static versus staggered arms & \texttt{P1/ER5} & \texttt{Q-REAL} \\
\attack{wire-eval} & Qualification & Mistral-7B; Mistral-NeMo-12B & Fixed obfuscation and per-epoch ratchet & Protected wire recovery falls while identity-restoration control remains correct & E9 baseline 50\% to obfuscated 0\%; rotated-epoch wire recovery 0\% in the ratchet tier & Unobfuscated baseline, shared-key restoration and RTT twins & \texttt{P1/E9}, \texttt{P1/ER} & \texttt{Q-REAL} \\
\attack{known-prefix} & Qualification & Qwen3-0.6B; Qwen/Mistral E-R tiers & Fixed or jittered prefix / rotation & More known prefix pairs cross the predeclared recovery fraction only when alignment survives & Qwen depth-1 $K_{90}$ approximately 232--240 for prefixes 32--128; non-crossings retained & Label-free band, prefix-length sweep and jittered arm & \texttt{P1/E8-KP}, \texttt{P1/ER3} & \texttt{Q-REAL} \\
\attack{alignment-search} & Qualification & Mistral-NeMo-12B E-R9 diagnostic & Prompt jitter and bidirectional folding & Realignment lifts recovery when jitter hides correspondence & Approximate alignment reached 15.21\% versus 9.71\% label-free, but no epoch produced an accepted solve & Unjittered scaffold, label-free band and solved-outcome gate & \texttt{P1/ER9} & \texttt{PARTIAL} \\
\attack{membership} & Qualification & Qwen3-0.6B & Split fine-tuning / FedAvg & Corrected seeded sampling yields repeatable membership AUC above 0.5 if features leak & Membership AUC 0.875$\pm$0.013 and 0.874$\pm$0.008 over three training seeds & Shuffled membership and distinct feature hashes & \texttt{P1/E-A4} & \texttt{Q-REAL} \\
\attack{view-privacy} & Qualification & GPT-2 plus exported bundle fixtures & Secret mixup, complementary views, tripwires & Oracle/coalition or repeated views expose more than a single protected view; honest controls remain usable & Bundle and repeated-view qualification completed; separate v3 attacks leave 83.01--90.67\% token recovery under mixup & Single view, known-mean oracle, honest correction and reset windows & \texttt{PA/V3}, \texttt{T/name} & \texttt{PARTIAL} \\
\attack{representation-privacy} & Qualification & GPT-2 plus bottleneck bundles & Trained bottleneck, PCA and Gram flattening & Fresh coordinate/norm/Gram probes detect residual signal; label shuffle remains null & Bundle/probe qualification completed; separate validated white-box attack recovers 18.85--68.59\% & Shuffled labels, fresh probes and public/private-phase controls & \texttt{PB/WB}, \texttt{T/name} & \texttt{PARTIAL} \\
\attack{agent-goal-hijack} & Qualification & Qwen3-0.6B planner + sandbox & Instruction provenance & Vulnerable simulator violates and defended simulator blocks without rejecting sanctioned actions & Synthetic direction reproduced; later live tier measured every setting with no violation but blocked legitimate actions in the sanctioned contrasts & Sanctioned request, no-control and provenance-only arms & \texttt{PB/AG}, \texttt{T/name} & \texttt{PARTIAL} \\
\attack{cross-agent-deputy} & Qualification & Qwen3-0.6B planner + agent bus & Principal binding versus message authentication & Authentic forwarded request violates without principal binding and is rejected with binding & Synthetic single-flag contrast reproduced; live planner produced no violation and false-blocked sanctioned requests in the measured tier & Sanctioned request and authentication-only distinctness arms & \texttt{PB/AG}, \texttt{T/name} & \texttt{PARTIAL} \\
\attack{gradient-hardening} & Qualification & GPT-2 124M & Gradient noise, clipping and 8/4-bit quantisation & True-gradient recovery remains above shuffled controls until hardening removes useful signal & At depth six/public, $\sigma{=}0.1$ gives 97.31\%, 8-bit gives 97.33\%, true gradient 97.38\%, shuffled 93.91\% & True, shuffled and random-candidate arms & \texttt{PA/V3} & \texttt{Q-REAL} \\

\midrule
\multicolumn{9}{@{}l}{\textbf{Synthetic and simulator qualification (35)}}\\
\midrule
\attack{leak-accumulation} & Synthetic & Generated epoch captures & Per-epoch row leakage / rotation & Persistent row leakage compounds across epochs rather than resetting & Seeded APT fixture produces the expected accumulation curve & Epoch-reset and zero-row-leak controls & \texttt{T/name} & \texttt{Q-SYN} \\
\attack{subspace-joint} & Synthetic & Generated shared-basis captures & Per-epoch keys & Joint basis solve needs fewer pairs than independent epoch solves when a basis is shared & Planted shared-basis fixture recovers; independent-basis null removes the advantage & Independent solves and no-shared-basis world & \texttt{T/name} & \texttt{Q-SYN} \\
\attack{seed-disclosure} & Synthetic & Generated encoder & Disclosed initialisation seed & Correct seed reproduces the encoder digest and lifts inversion; wrong seed does not & Digest and reconstruction checks pass in the fixture & Wrong/random seed and digest mismatch & \texttt{T/name} & \texttt{Q-SYN} \\
\attack{ica-bss} & Synthetic & Generated mixed sources & Unknown rotation / mixing & FastICA separates non-Gaussian sources above chance but not a null mixture & Deterministic ICA fixture qualifies the solver & Permuted-source and null-mixture controls & \texttt{T/name} & \texttt{Q-SYN} \\
\attack{active-cloud} & Synthetic & Split-model fixture & Crafted returned tensors / updates & Crafted cloud intervention increases leakage over passive honest behaviour & Planted intervention direction reproduced in the fixture & Honest cloud and zero-perturbation arms & \texttt{T/name} & \texttt{Q-SYN} \\
\attack{malicious-server} & Synthetic & Splitbench fixture & Crafted body computation & Crafted amplification makes returned gradients more decodable than honest gradients & Crafted arm exceeds the honest arm at the same fixture budget & Honest server and equal-budget passive arm & \texttt{T/name} & \texttt{Q-SYN} \\
\attack{gpu-memory} & Synthetic & Allocator / unified-memory fixture & Zero-on-free and isolation & Residue or contention canaries appear in the vulnerable world and disappear after scrubbing/isolation & Canary and contention directions reproduced & Scrubbed buffers and idle/isolation controls & \texttt{T/name} & \texttt{Q-SYN} \\
\attack{wire-shape} & Synthetic & Generated packet journal & Padding / shape hiding fixture & Message boundaries, phases and token counts are recoverable only from informative shapes & Clean fixture boundary F1 is 1.0 with low token-count error; null/padded arms bound interpretation & Empty, padded and shuffled-timing journals & \texttt{T/name} & \texttt{Q-SYN} \\
\attack{reasoning-replay} & Synthetic & Opaque reasoning-blob fixture & Session/model binding & Compatible replay succeeds only under matching provider/session assumptions & Compatibility and poisoned-blob arms execute & Incompatible model/session and answer-only controls & \texttt{T/name} & \texttt{Q-SYN} \\
\attack{protocol-integrity} & Synthetic & Split-protocol simulator & Replay, reorder, downgrade, tamper, oversize & Vulnerable endpoint accepts mutations; defended endpoint rejects them while accepting valid frames & All declared mutation classes reproduce the expected open/closed direction & Untampered valid-frame control & \texttt{T/name} & \texttt{Q-SYN} \\
\attack{key-session-abuse} & Synthetic & Key-lifecycle simulator & Nonce reuse, rollback, downgrade, seed retention & Vulnerable lifecycle exposes reuse/blast radius; bound fresh sessions do not & Per-arm abuse outcomes and clean session behaviour retained & Fresh nonce/session, current epoch and no-log controls & \texttt{T/name} & \texttt{Q-SYN} \\
\attack{resource-exhaustion} & Synthetic & Serving-stack simulator & Caps, queue fairness and cache limits & Adversarial workloads amplify cost in the vulnerable world; caps contain them without blocking benign load & Nine workload arms reproduce amplification/containment directions & Benign tenant and capped defended world & \texttt{T/name} & \texttt{Q-SYN} \\
\attack{poisoning-backdoor} & Synthetic & Softmax/RAG fixtures & Poison filtering and update checks & Attack success rises with poison budget; controls and defences lower it & Eight poison families execute across vulnerable and defended worlds & Zero poison and random corruption & \texttt{T/name} & \texttt{Q-SYN} \\
\attack{malicious-client} & Synthetic & Shared-body simulator & Isolation and input/update validation & Crafted clients extract or destabilise the vulnerable shared body; isolated world contains effects & Crafted, Sybil and instability arms reproduce their declared directions & Honest client and isolated tenant & \texttt{T/name} & \texttt{Q-SYN} \\
\attack{supply-chain} & Synthetic & Artifact registry fixture & Digest, provenance, version and safe-load gates & Permissive registry admits each mutation; enforcing registry rejects it before load & Substitution, tamper, unsigned promotion, rollback, drift and unsafe serialization all reproduce & Pristine signed registry loads with zero false rejection & \texttt{T/name} & \texttt{Q-SYN} \\
\attack{replica-integrity} & Synthetic & Replica fixture / in-repo channels & Median, K=2 abort and per-channel keys & One malicious replica is contained; collusion or shared-key compromise exposes the designed boundary & Single-fault, collusion, heterogeneity and disclosure arms execute & All-honest and independently keyed replicas & \texttt{T/name} & \texttt{Q-SYN} \\
\attack{multi-tenant-isolation} & Synthetic & Shared-serving simulator & Physical/session isolation & Vulnerable shared state leaks canaries; isolated control does not & Session, batch, cache, KV, error and allocator paths reproduce & Physically isolated negative control & \texttt{T/name} & \texttt{Q-SYN} \\
\attack{artifact-telemetry-leak} & Synthetic & Generated stores / logs & Redaction, permissions and retention & Planted secrets are found across encodings; redacted stores avoid false positives & Exact, base64, hex, truncated and token-split canaries and permission checks reproduce & Benign store, redacted canaries and retention deadline & \texttt{T/name} & \texttt{Q-SYN} \\
\attack{router-side-channel} & Synthetic & Generated MoE traces & Uniform or random router & Informative routing leaks properties/linkability; null router stays near chance & Property and linkability AUC $\geq0.9$ in the fixture; null near 0.5 & Uniform-router and flat-contention controls & \texttt{T/name} & \texttt{Q-SYN} \\
\attack{privacy-composition} & Synthetic & Seeded multi-surface artifacts & Independent / interacting leaks & Composition curve preserves marginals and reports synergy only when planted & First-failure budget and joint-versus-strongest metrics reproduce & Duplicate-input, independent and single-surface controls & \texttt{T/name} & \texttt{Q-SYN} \\
\attack{prompt-injection} & Synthetic & Application sandbox & Instruction provenance / policy layer & Injection causes an unauthorised effect only in the vulnerable world & Direct, indirect, encoded and cross-session directions reproduce & Benign prompt and defended policy world & \texttt{T/name} & \texttt{Q-SYN} \\
\attack{hidden-context-exposure} & Synthetic & Application-context fixture & Context isolation / output filtering & Vulnerable world reveals planted context; defended world suppresses it & Exact, semantic, prefix, encoded and multi-turn endpoints execute separately & No-secret and answer-only controls & \texttt{T/name} & \texttt{Q-SYN} \\
\attack{rag-vector-abuse} & Synthetic & Vector-store fixture & Tenant filters and provenance & Cross-tenant, poison or retention arms affect vulnerable retrieval but not enforced controls & Retrieval, membership, flooding and citation-substitution arms reproduce & Clean corpus, tenant filter and fresh-retention controls & \texttt{T/name} & \texttt{Q-SYN} \\
\attack{output-sink-injection} & Synthetic & Downstream sink fixture & Sink-specific policy layer & Unsafe output causes an effect only without the downstream gate & HTML, SQL, shell, template, URL, path, code and structured sinks reproduce & Benign output and defended sink & \texttt{T/name} & \texttt{Q-SYN} \\
\attack{misinformation-integrity} & Synthetic & Authoritative-corpus fixture & Grounding / citation checks & Fabricated or stale claims score worse than grounded answers & Groundedness, contradiction, calibration and decision-harm endpoints execute & Authoritative answer and no-evidence controls & \texttt{T/name} & \texttt{Q-SYN} \\
\attack{tool-misuse} & Synthetic & Agent sandbox & Tool scopes and budgets & Dangerous parameters/scope amplification execute only in the vulnerable world; authorised task remains usable & Violation, blocking and zero-false-block directions reproduce & Authorised task and single-control ablations & \texttt{T/name} & \texttt{Q-SYN} \\
\attack{identity-privilege-abuse} & Synthetic & Agent sandbox & Principal binding & Principal switch executes without binding and is rejected with binding & Confused-deputy and cross-session arms reproduce & Authorised same-principal request & \texttt{T/name} & \texttt{Q-SYN} \\
\attack{agentic-supply-chain} & Synthetic & Agent or component registry & Component provenance & Tampered connector redirects the action only without digest-bound provenance & Bad-digest and trusted-label distinctness cases reproduce & Valid component and provenance-only arm & \texttt{T/name} & \texttt{Q-SYN} \\
\attack{unexpected-code-execution} & Synthetic & Canned-execution sandbox & Digest/task-bound execution gate & Unapproved digest executes only in the vulnerable world; approved digest remains usable & Direct, document and approval-swap arms reproduce & Authorised execution and single-flag ablations & \texttt{T/name} & \texttt{Q-SYN} \\
\attack{memory-context-poisoning} & Synthetic & Persistent-memory sandbox & Memory provenance & Cross-session planted directive changes action only without provenance & Poisoned-memory direction and defended block reproduce & Clean memory and authorised persistence & \texttt{T/name} & \texttt{Q-SYN} \\
\attack{insecure-agent-communication} & Synthetic & In-memory agent bus & Message authentication / nonce freshness & Spoof, tamper and replay execute only without endpoint verification & Three communication attacks and seven distinctness worlds reproduce & Valid signed message and fresh nonce & \texttt{T/name} & \texttt{Q-SYN} \\
\attack{cascading-failure} & Synthetic & Multi-hop agent pipeline & Per-hop revalidation & Broken propagation chain acts only without revalidation; intact authorised chain remains usable & Broken/intact and owner/attacker distinctness cases reproduce & Intact authorised chain and revalidation-only arm & \texttt{T/name} & \texttt{Q-SYN} \\
\attack{human-agent-trust} & Synthetic & Human-approval fixture & Faithful display & Displayed/actual mismatch executes only without display integrity & Mismatch and faithful-display arms reproduce & Faithful summary and authorised refund & \texttt{T/name} & \texttt{Q-SYN} \\
\attack{rogue-agent} & Synthetic & Lifecycle sandbox & Lifecycle attestation & Unattested replication, self-modification or persistence executes only without attestation & All three lifecycle classes and authorised spawn distinctness reproduce & Operator-attested lifecycle action & \texttt{T/name} & \texttt{Q-SYN} \\
\attack{two-seam-delegation} & Synthetic & Generated two-seam model & Private middle segment & Chained decode beats direct seam-2 decode when the middle is planted; identity middle collapses to the single-seam control & Planted and identity worlds reproduce, including noise arms & Single-seam and identity-middle controls & \texttt{T/name} & \texttt{Q-SYN} \\

\midrule
\multicolumn{9}{@{}l}{\textbf{Orchestration helpers (3)}}\\
\midrule
\attack{remote-fetch} & Unmeasured / helper & Configured fleet & Reachability and access controls & Reachable captures are copied with provenance; unreachable/disabled nodes are recorded cleanly & Orchestration behaviour and clean unreachable-node handling are tested & Disabled and unreachable node controls & \texttt{T/name} & \texttt{HELPER} \\
\attack{mitm-proxy} & Unmeasured / helper & Local relay fixture & Transparent forwarding & Forwarded bytes remain unchanged while both directions are journalled & Relay/tee behaviour and byte identity are tested & Direct endpoint and byte-identity check & \texttt{T/name} & \texttt{HELPER} \\
\attack{max-effort} & Unmeasured / helper & Wrapped attack & Budget / seed sweep & Wrapper preserves the target attack and reports its distribution over effort & Interface and effort-distribution machinery are qualified & Single-budget wrapped run & \texttt{T/name} & \texttt{HELPER} \\
\end{longtable}
\end{landscape}

\paragraph{Full-precision numbers.}
The main experiment's primary difference is $3.1714$ percentage points; the
Wilson multiplier is $z = 2.5391848136513127$; the bootstrap uses 100{,}000
replicates with seed 93281. The domain-battery floors round to 3.911, 2.603
and 2.342 percentage points. The main experiment used 336.1 hours of summed
online attack time and wrote 17 hash-chained logs. The validated white-box
runs each record 128 optimisation steps at learning rate 0.05, a victim
learning rate of $5\times10^{-5}$, 64 training steps, 32 calibration and 32
attacked sequences, 1{,}024 scored positions, 131{,}072 token positions in
each direction and 51{,}463{,}168 vocabulary comparisons in readout, with
1{,}223.24 seconds of attack time over the 198 stable runs and 617.16 seconds
over the 99 co-trained runs. The BiSR implementation check trained six
inverters for 20 epochs on 112 auxiliary articles (15{,}604 windows), ran 24
prediction settings (6{,}144 document-engine runs) and used an equivalence
interval of $(-0.01, 0.01)$ with $t = 2.919985580355516$; every one of the
32 comparisons gave the interval $[0,0]$, which describes the agreement on
these fixed documents. The adapted comparison ran two attack seeds (17, 29) in 80 joint
executions, for 160 scored runs and 20{,}480 document executions, reused 240
control runs, trained four GRU inverters, one per depth and attack seed, for
twenty epochs, applied 32
gradient-matching and 800 activation-matching steps, multiplied the stored
gradients by 256 to convert the loss normalisation to a per-document mean,
and used 29.70 hours of prediction time and 0.58 hours of preparation.
Independent audits on two CPU hosts of the 500-step extension verified all
13 execution journals and 12{,}443 raw files. The live-planner canaries and
full run used 234.79 seconds of worker wall time.

\paragraph{Synthetic order-and-decoy battery.}
Some defences shuffle the order of the activations and add decoy rows. A
synthetic battery, with data generated by the framework, ran the matching
module against such data for three seeds. In its six matching reports,
the highest excess of order recovery over chance ranged from 0.52 to 5.21
points. The matching module is a local supervised baseline, and the battery
shows that the framework scores the order of the
rows separately from their tokens.

\clearpage
\begingroup
\small
\bibliographystyle{plain}
\bibliography{main}
\endgroup
\end{document}